\documentstyle[aps,psfig,epsfig]{revtex}

\newcommand{\pr}{\prime}
\newcommand{\na}{\nabla}

\newcommand{\ep}{\epsilon}

\newcommand{\vphi}{\varphi}

\newcommand{\lraw}{\longrightarrow}

\newcommand{\pa}{\partial}

\newcommand{\td}{\tilde}
\newcommand{\sla}[1]{\slash\!\!\! #1}

\begin{document}

\draft
\title{Chiral Expansion Theory at Vector Meson Scale}
\author{Xiao-Jun Wang\footnote{E-mail address: wangxj@mail.ustc.edu.cn}}
\address{Center for Fundamental Physics,
University of Science and Technology of China\\
Hefei, Anhui 230026, P.R. China}
\author{Mu-Lin Yan\footnote{E-mail address: mlyan@staff.ustc.edu.cn}}
\address{CCST(World Lad), P.O. Box 8730, Beijing, 100080, P.R. China \\
  and\\
 Center for Fundamental Physics,
University of Science and Technology of China\\
Hefei, Anhui 230026, P.R. China\footnote{mail
address}}
\date{\today}
\maketitle

\begin{abstract}
We study physics on $\rho(770)$ and $\omega(782)$ in framework of chiral
constituent quark model. The effective action is derived by proper vertex
method, which can capture all order information of chiral expansion. The
$N_c^{-1}$ expansion is also studied systematically. It is shown that the
momentum expansion at vector meson energy scale converges slowly, and the
loop effects of pseudoscalar meson play an important role at this energy
scale. We provide a method to prove the unitarity of S-matrix in
any low-energy effective theory of QCD. Phenomenologically, we study
decays for $\rho\rightarrow\pi\pi$, $\rho\rightarrow e^+e^-$,
$\omega\rightarrow\pi^+\pi^-$, $\rho^\pm\rightarrow\gamma\pi^\pm$ and
$\omega\rightarrow\gamma\pi^0$, $\rho^0-\omega$ mixing and their mass
splitting, pion form factor, $I=l=1$ phase shift and light quark masses at
vector meson energy scale. These results include all order contribution of
vector meson momentum expansion and is up to next to leading order of
$N_c^{-1}$ expansion. All of these theoretical predictions agree with
data very well. The unitarity of S-matrix yielded by this framework is
examined. The Breit-Winger formula for resonance propagator is derived.
\end{abstract}

\section{Introduction}
In past thirty years, the effective theory studies on physics of
light vector meson resonances($\rho$, $\omega$, $K^*$ and $\phi$)
have attracted much interests. Works in this field have helped us
to understand various aspects on low energy QCD and properties of
light vector mesons. A challenging subject in these studies is to
construct a well-defined chiral expansion theory at vector meson
scale. In ref.\cite{Manohar95} authors have developed a chiral
expansion theory including vector mesons. In this framework one
can successfully deal with two-vector-meson processes (one vector
meson is in initial state and one is in final state)\cite{Appl}.
However, it was also pointed out in ref.\cite{Manohar95} that
this framework is failed to study single-vector-meson processes.
So far, the most studies on single-vector-meson processes are
still limited at the lowest order of vector meson four-momentum
expansion and $N_c^{-1}$ expansion. Theoretical reason is that
convergence of momentum expansion in these processes is unclear,
and technical reason is that the calculation beyond the leading
order is very difficult. For example, method of chiral
perturbative theory(ChPT)\cite{GL85a} is impractical for these
processes because more and more free parameters are required as
perturbative order raising. Obviously, this is a serious problem
that has to be resolved, since it makes the models' calculations
being not controlled approximations in that error bars can not be
put on the predictions. Although there are some attempts to resolve these
problem in terms of resumming the momentum expansion at vector
meson scale\cite{resume}, the resummation was not yet used in
phenomenological studies of vector mesons so far. The purposes of
this present paper are to study chiral expansion in
single-vector-meson processes systematically in the framework of
chiral quark model, and to overcome the difficulties mentioned in
above. Our studies will be not only resumming momentum expansion
of vector mesons, but also beyond the leading order of $N_c^{-1}$
expansion.

The convergence of the chiral expansion is a fundamental
requirement for any chiral model. In this paper, a broad class of
chiral quark model will be checked by this requirement. The
simplest version of the chiral quark model is in ref.\cite{MG84},
that was originated by Weinberg\cite{Wein79}, and has been
extended to some different versions\cite{Esp90,Chan85,ENJL,Bijnens93}. The
constituent quark model in ref.\cite{MG84} is different from
other versions in refs.\cite{Esp90,Chan85,ENJL,Bijnens93}. The massless
pseudoscalar mesons are treated as fundamental dynamical field
degrees of freedom in the former, but are treated as composite
fields of quarks in the latter. In section 5 of this paper, we
will show that the chiral quark models in
refs.\cite{Esp90,Chan85,ENJL,Bijnens93} can not yield convergent chiral
expansion in single-vector-meson processes, and the effective
lagrangian derived from these chiral quark models can not give
unitarity S-matrix for vector meson physics. Therefore, in this
paper, our studies will be within the framework of a chiral
constituent quarks model proposed by Manohar and
Georgi\cite{MG84}(quote as MG model hereafter). It is the
simplest and practical model for achieving our goal. In view of
this model, in energy region between chiral symmetry spontaneously
breaking(CSSB) scale($\Lambda_\chi\simeq 2\pi f_\pi\simeq
1.16$GeV) and the confinement scale($\Lambda_{QCD}\sim
0.1-0.3$GeV), the dynamical field degrees of freedom are
constituent quarks(quasi-particle of quarks), gluons and
Goldstone bosons associated with CSSB(it is well-known that these
Goldstone bosons correspond to lowest pseudoscalar octet). In this
quasiparticle description, the effective coupling between gluon
and quarks is small and the important interaction is the coupling
between quarks and Goldstone bosons. The effective action
describing interaction of Goldstone bosons can be obtained via
integrating out freedoms of quark fields.

The MG model and its extension have been studied continually
during the last fifteen years and got great success in different
aspects of phenomenological predictions in hadron
physics\cite{EMG}. Low energy limit of the model has been
examined successfully\cite{Esp90,Wang98}. In this present paper,
the MG model will be extended to include lowest vector meson
resonances(quote as extended Manohar-Georgi model or EMG model
hereafter) in terms of WCCWZ realization for spin-1
mesons\cite{Wein68,WCCWZ}. The advantages of approach of the MG
model are that, in principle, high order contribution of the
chiral expansion can be computed completely, and only fewer free
parameters are required. So far, however, there are still no any
systematic studies beyond the lowest order of the chiral
expansion due to technical difficulties. After freedoms of
fermions are integrated out, traditionally, the determinant of
Dirac operator are regularized by some standard methods, such as
Schwinger proper time method\cite{Sch54}, or heat kernel
method\cite{Ball89}, and are expanded in powers of momentum of
mesons. In terms of this method, the effective action can keep
chiral symmetry explicitly order by order. However, it is very
difficult to capture arbitrary order contribution in chiral
expansion by means of the above method. In this present paper, we
will propose another equivalent method to evaluate the effective
action via calculating one-loop diagrams of constituent quarks
directly. We call it as proper vertex expansion method. By this
method, the effective action is expanded in numbers of external
vertices instead of external momentum. The effective action with
n external vertices is just connective n-point Green function of
EMG model in presence of external fields. It includes all order
contribution of the momentum expansion. Thus in terms of this
method, the chiral theory at vector meson energy scale can be
studies up to all orders of the momentum expansion.

Another challenging subject is to study vector meson physics
beyond large $N_c$ limit. In the other words, how can we
calculate meson loops systematically and consistently? At vector
meson energy scale, it can be expected that the contributions
from pseudoscalar meson loops play important and highly nontrivial
role. This is due to the following reasons: 1)Empirically, for
vector meson physics, the contribution from pseudoscalar meson
loops is about $\Gamma_\rho:m_\rho\simeq 20\%$. The conclusion is
obtained because the width of $\rho$-meson is generated
dynamically by pion loops and $\Gamma_\rho:m_\rho$ is suppressed
by $N_c^{-1}$ expansion according to standard $N_c^{-1}$
argument\cite{tH74}. It indicates that the contribution from
pseudoscalar meson loops is not small so that it can not be
omitted simply. 2) Since both pseudoscalar mesons and constituent
quarks are fundamental dynamical field degrees of freedoms at this
energy scale, theoretically, loop effects of pseudoscalar mesons
should be also  considered as well when effective action is
derived by loop effects of constituent quarks. 3)For physical
value $N_c=3$, the $N_c^{-1}$ expansion is not convergent very
rapidly. Due to the above reasons, in this paper we will also
study contribution from pseudoscalar meson loops systematically.
The calculations on meson loops will suffer ultraviolet(UV)
divergent difficulty. At vector meson energy scale, loop effects
of pseudoscalar mesons cause $\phi-\omega$ mixing which will
destroy OZI rule\cite{OZI} if this contribution is not to be an
appropriate value. In the other words, it can insure OZI rule for
$\phi$ to choose an UV cut-off appropriately. Thus, very
fortunately, the OZI rule for $\phi$ provides a natural way to
determine all UV-dependent contributions in the meson loop
calculations.

The unitarity of S-matrix is a fundamental requirement for any quantum
field theory. It is well-known that the unitarity of S-matrix of an
effective theory can not be insured in full energy region, but only be 
insured in energy region below an energy cut-off. The consistency of this
effective theory requires masses of all resonances should be smaller than
this cut-off. In the other words, it indicates that the cut-off
$\Lambda_{\rm EMG}$ in EMG model should satisify $\Lambda_{\rm
EMG}>m_{\omega}$ for SU(2) and $\Lambda_{\rm EMG}>m_{\phi}$ for SU(3) at
least. In general, a low energy effective field theory of QCD is a
well-defined perturbative theory in $N_c^{-1}$ expansion. Thus the
unitarity of S-matrix of low energy effective field theory has to be
satisfied in order by order in $N_c^{-1}$ expansion. For EMG model, the
unitarity of S-matrix will be examined explicitly in this paper. We will
see that the meson-loop effects play an essential role for meeting the
requirement of the unitarity.

The phenomenology of vector meson physics is very rich. For
example, the vector meson dominance(VMD) and universal coupling
for vector mesons, which are good approximation for vector meson
physics\cite{Current}, should be satisfied in the leading order of
the chiral expansion. In this present paper, we will achieve
various phenomenological results: 1)The decays for
$\rho\rightarrow\pi\pi$, $\rho\rightarrow e^+e^-$. 2) Dynamical
width for $\rho$-meson and Breit-Wigner formula for propagator of
resonance. 3) $\rho^0-\omega$ mixing and
$\omega\rightarrow\pi^+\pi^-$ decay. The mass difference between
$\rho$ and $\omega$ are also predicted. We will show that, for
resonance particles with large width(e.g., $\rho$), the mass
parameter in its propagator is not just the its physics mass.
4)Pion form factor and $I=1,\;l=1$ pion-pion phase shift. Some
nontrivial phase in pion form factor are revealed too. 5)
Individual light current quark masses at vector meson energy
scale. The quark masses are model-dependent and are necessary to
extend our study to $K^*$ and $\phi$ mesons. 6)The anomalous
decays $\rho^\pm\rightarrow\gamma\pi^\pm$ and
$\omega\rightarrow\gamma\pi$. All above results resum all
information of the momentum expansion of vector mesons and are up to the
next to leading order of $N_c^{-1}$ expansion.

The paper is organized as follows. In sect. 2 the Manohar-Georgi
model and its extension are reviewed. In sect. 3, the relevant
effective action is derived at the leading order of $N_c^{-1}$
expansion. This effective action includes all order contribution
of the chiral expansion in powers of vector meson four-momentum
square. The calculation on pseudoscalar meson loops will be
included in the following sections. In sect. 4, decays for
$\rho\rightarrow\pi\pi$ and $\rho^0\rightarrow e^+e^-$ are
studied. In sect. 5, the unitarity of S-matrix is examined
explicitly and Breit-Wigner formula for $\rho$-propagator is
obtained. In sect. 6, the $\rho^0-\omega$ mixing and
$\omega\rightarrow\pi\pi$ decay are studied. The isospin breaking
parameter $m_d-m_u$ is predicted at energy scale $\mu=m_\rho$. In
sect. 7, the anomalous decays for
$\rho^\pm\rightarrow\gamma\pi^\pm$ and
$\omega\rightarrow\gamma\pi$ are predicted. In sect. 8, pion form
factor, $I=1,\;l=1$ $\pi-\pi$ phase shift and $\rho^0-\omega$
mass difference are studied. The light quark masses are predicted
in sect. 9 and a brief summary is in sect. 10.

\section{Manohar-Georgi model and its extension}

For interpreting physics below CSSB scale, Manohar and Georgi provides a
QCD-inspired description on the simple constituent quark model. At chiral
limit, it is parameterized by the following $SU(3)_{_V}$ invariant chiral
constituent quark lagrangian
\begin{eqnarray}\label{2.1}
{\cal L}_{\chi}&=&i\bar{\psi}(\sla{\pa}+\sla{\Gamma}+
  g_{_A}{\slash\!\!\!\!\Delta}\gamma_5)\psi-m\bar{\psi}\psi
   +\frac{F^2}{16}Tr_f\{\nabla_\mu U\nabla^\mu U^{\dag}\}.
\end{eqnarray}
Here $Tr_f$ denotes trace in SU(3) flavour space,
$\bar{\psi}=(\bar{u},\bar{d},\bar{s})$ are constituent quark fields,
$g_{_A}=0.75$ can be fitted by beta decay of neutron. The $\Delta_\mu$
and $\Gamma_\mu$ are defined as follows,
\begin{eqnarray}\label{2.2}
\Delta_\mu&=&\frac{1}{2}\{\xi^{\dag}(\pa_\mu-ir_\mu)\xi
          -\xi(\pa_\mu-il_\mu)\xi^{\dag}\}, \nonumber \\
\Gamma_\mu&=&\frac{1}{2}\{\xi^{\dag}(\pa_\mu-ir_\mu)\xi
          +\xi(\pa_\mu-il_\mu)\xi^{\dag}\},
\end{eqnarray}
and covariant derivative are defined as follows
\begin{eqnarray}\label{2.3}
\nabla_\mu U&=&\pa_\mu U-ir_\mu U+iUl_\mu=2\xi\Delta_\mu\xi,
  \nonumber \\
\nabla_\mu U^{\dag}&=&\pa_\mu U^{\dag}-il_\mu U^{\dag}+iU^{\dag}r_\mu
  =-2\xi^{\dag}\Delta\xi^{\dag},
\end{eqnarray}
where $l_\mu=v_\mu+a_\mu$ and $r_\mu=v_\mu-a_\mu$ are linear
combinations of external vector field $v_\mu$ and axial-vector
field $a_\mu$, $\xi$ associates with non-linear realization of
spontaneously broken global chiral symmetry $G=SU(3)_L\times
SU(3)_R$ introduced by Weinberg\cite{Wein68},
\begin{equation}\label{2.4}
\xi(\Phi)\rightarrow
g_R\xi(\Phi)h^{\dag}(\Phi)=h(\Phi)\xi(\Phi)g_L^{\dag},\hspace{0.5in}
 g_L, g_R\in G,\;\;h(\Phi)\in H=SU(3)_{_V}.
\end{equation}
Explicit form of $\xi(\Phi)$ is usual taken
\begin{equation}\label{2.5}
\xi(\Phi)=\exp{\{i\lambda^a \Phi^a(x)/2\}},\hspace{1in}
U(\Phi)=\xi^2(\Phi),
\end{equation}
where the Goldstone boson $\Phi^a$ are treated as pseudoscalar meson
octet. The constituent quark fields transform as matter fields of
SU(3)$_{_V}$,
\begin{equation}\label{2.6}
  \psi\lraw h(\Phi)\psi, \hspace{1in} \bar{\psi}\lraw
\bar{\psi}h^{\dag}(\Phi).
\end{equation}
$\Delta_\mu$ is SU(3)$_{_V}$ invariant field gradients and $\Gamma_\mu$
transforms as field connection of SU(3)$_{_V}$
\begin{equation}\label{2.7}
\Delta_\mu\lraw h(\Phi)\Delta_\mu h^{\dag}(\Phi), \hspace{0.8in}
\Gamma_\mu\lraw h(\Phi)\Gamma_\mu h^{\dag}(\Phi)+h(\Phi)\pa_\mu
  h^{\dag}(\Phi).
\end{equation}
Thus the lagrangian(~\ref{2.1}) is invariant under $G_{\rm global}\times
G_{\rm local}$.

For achieving the purposes of this present paper, MG model must
be extended to including lowest vector meson resonance and beyond
chiral limit. The light quark mass matrix ${\cal M}={\rm
diag}\{m_u,m_d,m_s\}$ is usually included into external spin-0
fields, i.e., $\td{\chi}=s+ip$, where $s=s_{\rm ext}+{\cal M}$,
$s_{\rm ext}$ and $p$ are scalar and pseudoscalar external fields
respectively. The chiral transformation for $\td{\chi}$ is
\begin{eqnarray}\label{2.18}
\td{\chi}\rightarrow g_{_R}\td{\chi}g_{_L}^{\dag}.
\end{eqnarray}
Thus $\td{\chi}$ and $\td{\chi}^{\dag}$ together with $\xi$ and
$\xi^{\dag}$ can form SU(3)$_{V}$ invariant scalar source
$\xi^{\dag}\td{\chi}\xi^{\dag}+\xi\td{\chi}^{\dag}\xi$ pseudoscalar
source $(\xi^{\dag}\td{\chi}\xi^{\dag}-\xi\td{\chi}^{\dag}\xi)\gamma_5$.
Then current quark mass dependent term is written
\begin{equation}\label{2.19}
-\frac{1}{2}\bar{q}(\xi^{\dag}\td{\chi}\xi^{\dag}
+\xi\td{\chi}^{\dag}\xi)q-\frac{\kappa}{2}
 \bar{q}(\xi^{\dag}\td{\chi}\xi^{\dag}-\xi\td{\chi}^{\dag}\xi)
 \gamma_5q.
\end{equation}
Eq.~(\ref{2.19}) will return to standard quark mass term of QCD
lagrangian, $\bar{\psi}{\cal M}\psi$, in absence of pseudoscalar mesons at
high energy for arbitrary $\kappa$. It means that the symmetry and some
underlying constrains of QCD can not fixed the coupling between
pseudoscalar mesons and constituent quarks. Hence $\kappa$ is treated as
an initial parameter of the model and will be fitted phenomenologically.

From the viewpoint of chiral symmetry only, an alternative scheme for
incorporating vector mesons was suggested by Weinberg\cite{Wein68} and
developed by Callan, Coleman et. al\cite{WCCWZ}. In this treatment, vector
meson resonances $V_\mu$ transform homogeneously under SU(3)$_{V}$,
\begin{eqnarray}\label{2.20}
V_\mu\rightarrow h(\Phi)V_\mu h^{\dag}(\Phi),
\end{eqnarray}
where
\begin{equation}\label{2.21}
   V_\mu(x)={\bf \lambda \cdot V}_\mu =\sqrt{2}
\left(\begin{array}{ccc}
       \frac{\rho^0_\mu}{\sqrt{2}}+\frac{\omega_\mu}{\sqrt{2}}
            &\rho^+_\mu &K^{*+}_\mu   \\
    \rho^-_\mu&-\frac{\rho^0_\mu}{\sqrt{2}}+\frac{\omega_\mu}{\sqrt{2}}
            &K^{*0}_\mu   \\
       K^{*-}_\mu&\bar{K}^{*0}_\mu&\phi_\mu
       \end{array} \right).
\end{equation}
Then extended Manohar-Georgi model(EMG model) is parametered by the
following SU(3)$_{_V}$ invariant lagrangian
\begin{eqnarray}\label{2.22}
{\cal L}_{\chi}&=&i\bar{q}(\sla{\pa}+\sla{\Gamma}+
  g_{_A}{\slash\!\!\!\!\Delta}\gamma_5-i\sla{V})q-m\bar{q}q
  -\frac{1}{2}\bar{q}(\xi^{\dag}\td{\chi}\xi^{\dag}
  +\xi\td{\chi}^{\dag}\xi)q
  -\frac{\kappa}{2}\bar{q}(\xi^{\dag}\td{\chi}\xi^{\dag}
 -\xi\td{\chi}^{\dag}\xi)\gamma_5q \nonumber \\
   &&+\frac{F^2}{16}Tr_f\{\nabla_\mu U\nabla^\mu U^{\dag}\}
   +\frac{1}{4}m_0^2Tr_f\{V_\mu V^{\mu}\}.
\end{eqnarray}
We can see that there are five initial parameters
$g_{_A},\;m,\;\kappa,\;F$ and $m_0$ in EMG model. These parameters can
not be determined by symmetry and should be fitted by experiment. The
effective lagrangian describing interaction of vector meson resonances
will be generated via loop effects of constituent quarks.

It should carefully distinguish EMG model from ENJL-like
models\cite{Chan85,ENJL,AE99} which seem to be similar to EMG model. The
difference is that there are kinetic term of pseudoscalar mesons in EMG
model, but this term is absent in ENJL-like models. It makes there is a
extra constrain on constituent quark mass $m$ in ENJL-like models, and
this constrain requires constituent quark mass $m$ around 300MeV. However,
in sect. V we will point out that this value of $m$ can not insure
convergence of chiral expansion in single-vector-meson processes.

\section{Effective action and proper vertex expansion}
\setcounter{equation}{0}

\subsection{Proper vertex expansion}

The essential subject of nonperturbative QCD is to obtain a low
energy effective action of light hadrons, which belongs to the
leading order of $N_c^{-1}$ expansion. In chiral quark model,
this effective action is generated via loop effects of quarks,
i.e., integrating over degrees of freedom of quark. This path
integral can be performed explicitly. The determiant of fermion
obtained by path integral are usually regularized by heat kernel
method. In this method, the effective action is expanded in
powers of momentum of mesons. It is very difficult to use this
method to calculate high order contribution of momentum
expansion. In this present paper, we will derive the effective
action via computing the feymann diagrams of quark loops directly
instead of performing path integral.

We start with constituent quark lagrangian~(\ref{2.22}), and define vector
auxiliary field $\bar{V}_\mu^a(a=0,1,\cdots,8)$, axial-vector
auxiliary field $\Delta_\mu^a$, scalar auxiliary field
$S^a$ and pseudoscalar auxiliary field $P^a$ as follows
\begin{eqnarray}\label{3.1}
\bar{V}_\mu^\alpha&=&\frac{1}{2}Tr_f\{\lambda^\alpha(V_\mu+i\Gamma_\mu)\},
 \hspace{0.8in}
\Delta_\mu^a=\frac{1}{2}Tr_f\{\lambda^a\Delta_\mu\},\nonumber \\
S^a&=&\frac{1}{4}Tr_f\{\lambda^a(\xi^{\dag}\td{\chi}\xi^{\dag}
  +\xi\td{\chi}^{\dag}\xi)\}, \hspace{0.55in}
P^a=\frac{\kappa}{4}Tr_f\{\lambda^a(\xi^{\dag}\td{\chi}\xi^{\dag}
  -\xi\td{\chi}^{\dag}\xi)\},
\end{eqnarray}
where $\lambda^1,\cdots,\lambda^8$ are SU(3) Gell-Mann matrices and
$\lambda^0=\sqrt{\frac{2}{3}}$. Then in lagrangian~(\ref{2.22}), the
terms associating with constituent quark fields can be rewritten as follow
\begin{eqnarray}\label{3.2}
{\cal L}_\chi^q=\bar{q}(i\sla{\pa}-m)q
  +\bar{V}_\mu^a\bar{q}\lambda^a\gamma^\mu q
  +ig_A\Delta_\mu^a\bar{q}\lambda^a\gamma^\mu\gamma_5q
  -S^a\bar{q}\lambda^aq-P^a\bar{q}\lambda^a\gamma_5q.
\end{eqnarray}

The effective action describing meson interaction can be obtained via
loop effects of constituent quarks
\begin{eqnarray}\label{3.4}
e^{iS_{\rm eff}}&=&<0|T_qe^{i\int d^4x{\cal L}^{\rm I}_\chi(x)}|0>
       \nonumber \\
 &=&\sum_{n=1}^\infty i\int d^4p_1\frac{d^4p_2}{(2\pi)^4}
  \cdots\frac{d^4p_n}{(2\pi)^4}\tilde{\Pi}_n(p_1,\cdots,p_n)
  \delta^4(p_1-p_2-\cdots-p_n) \nonumber \\
&\equiv&i\Pi_1(0)+\sum_{n=2}^\infty i\int \frac{d^4p_1}{(2\pi)^4}
  \cdots\frac{d^4p_{n-1}}{(2\pi)^4}\Pi_n(p_1,\cdots,p_{n-1}),
\end{eqnarray}
where $T_q$ is time-order product of constituent quark fields,
${\cal L}_{\chi}^{\rm I}$ is interaction part of lagrangian~(\ref{3.2}),
$\tilde{\Pi}_n(p_1,\cdots,p_n)$ is one-loop effects of constituent quarks
with $n$ auxiliary fields, $p_1,p_2,\cdots,p_n$ are four-momentas of $n$
auxiliary fields respectively and
\begin{equation}\label{3.5}
\Pi_n(p_1,\cdots,p_{n-1})=\int d^4p_n\tilde{\Pi}_n(p_1,\cdots,p_n)
  \delta^4(p_1-p_2-\cdots-p_n).
\end{equation}
To get rid of all disconnected diagrams, we have
\begin{eqnarray}\label{3.6}
S_{\rm eff}&=&\sum_{n=1}^\infty S_n, \nonumber \\
S_1&=&\Pi_1(0),  \\
S_n&=&\int \frac{d^4p_1}{(2\pi)^4}\cdots\frac{d^4p_{n-1}}
  {(2\pi)^4}\Pi_n(p_1,\cdots,p_{n-1}), \hspace{0.8in}(n\geq 2)\nonumber.
\end{eqnarray}
Obviously, in eq.~(\ref{3.6}) the effective action $S_{\rm eff}$
is expanded in powers of number of external vertex and is
expressed as integral over external momentum. Hereafter we will
call this method as proper vertex method, and call $S_n$ as
$n$-point effective action.

\subsection{Relevant effective action}

In terms of proper vertex method, the effective action can be
obtained, and in principle it can include all information on
resumming momentum expansion. Practically, however, it is tedious
to express the effective action completely. In this section, we
only derive those effective action relating to phenomenological
studies on $\rho$ and $\omega$ physics in next sections. For
simplifying the expression of the effective action, some good
approximation will be used. Since we work in $\rho$-meson energy
scale, the expansion of light quark masses still works very well.
We can set $m_u=m_d=0$ in $\rho$ physics and take leading order of
$m_d-m_u$ expansion in isospin breaking $\omega$ physics. This
also implies that the soft-pion theorem for pseudoscalar mesons
is available here.

The one-point effective action $S_1$ is just contribution of
tadpole diagram of constituent quark,
\begin{eqnarray}\label{3.12}
S_1=\int d^4x\frac{F_\pi^2}{8}B_0
 Tr_f\{\td{\chi} U^{\dag}+\td{\chi}^{\dag}U\},
\end{eqnarray}
where $B_0$ is a constant which absorb the quadratic divergence from quark
loop integral
\begin{eqnarray}\label{3.13}
\frac{F_0^2}{8}B_0=\frac{N_c}{(4\pi)^{D/2}}(\frac{\mu^2}{m^2})^{\ep/2}
   \Gamma(1-\frac{D}{2})m^3.
\end{eqnarray}

After renormalizing the kinetic term of pseudoscalar mesons, the two-point
effective action reads
\begin{eqnarray}\label{3.14}
S_2&=&\frac{F_\pi^2}{16}\int d^4xTr_f\{\na_\mu U\na^\mu U^{\dag}\}
    +\frac{1}{4}m_0^2\int d^4xTr_f\{V_\mu(x)V^\mu(x)\}\nonumber\\
&&-\int\frac{d^4p}{(2\pi)^4}\frac{\alpha_1(p^2)}{4}
  (g_{\mu\nu}p^2-p_\mu p_\nu)Tr_f\{(V^\mu(p)+i\Gamma^\mu(p))
  (V^\nu(-p)+i\Gamma^\nu(-p))\}+...,
\end{eqnarray}
where ``...'' denotes those terms are independent of the following
phenomenological studies of this present paper, and
\begin{eqnarray}\label{3.15}
\alpha_1(p^2)&=&g^2-\frac{N_c}{\pi^2}\int_0^1\cdot x(1-x)
  \ln(1-\frac{x(1-x)p^2}{m^2}),\nonumber \\
{\cal O}(p)&=&\int d^4x e^{ip\cdot x}{\cal O}(x),\hspace{1in}
{\cal O}=V,\;\Gamma,\;\Delta,\cdots.
\end{eqnarray}
The constant $g$ in eq.~(\ref{3.15}) is an universal coupling
constant, which absorbs the logarithmic divergence from quark
loop integral,
\begin{eqnarray}\label{3.16}
\frac{3}{8}g^2=\frac{N_c}{(4\pi)^{D/2}}(\frac{\mu^2}{m^2})^{\ep/2}
  \Gamma(2-\frac{D}{2}).
\end{eqnarray}

It is obvious that the second term of $S_2$ includes all order information
of momentum expansion. In coordinate space, it can be expanded in powers
of number of derivative,
\begin{eqnarray}\label{3.17}
-\frac{g^2}{8}Tr_f\{\bar{V}_{\mu\nu}\bar{V}^{\mu\nu}\}
 -\frac{N_c}{240\pi^2}m^{-2}Tr_f\{\pa_\rho\bar{V}_{\mu\nu}
  \pa^\rho\bar{V}^{\mu\nu}\}
-\frac{N_c}{2240\pi^2}m^{-4}Tr_f\{\pa_\rho\pa_\sigma\bar{V}_{\mu\nu}
  \pa^\rho\pa^\sigma\bar{V}^{\mu\nu}\}-\cdots,
\end{eqnarray}
with
\begin{eqnarray}\label{3.18}
\bar{V}_{\mu\nu}=\pa_\mu\bar{V}_\nu-\pa_\nu\bar{V}_\mu,
\hspace{1in}\bar{V}_\mu=V_\mu+i\Gamma_\mu.
\end{eqnarray}
In next section we will show that the high order derivative terms yield
very important contribution at vector meson energy scale, i.e., for
$p^2\sim m_\rho^2$.

The three-point effective action includes two parts. One describes
reactions with normal parity and another one describes reactions with
abnormal parity. For the purpose of this present paper, the three-point
effective action with normal parity can be obtained via standard
calculation of feynman diagram(fig.1),
\begin{figure}[hptb]
   \centerline{\psfig{figure=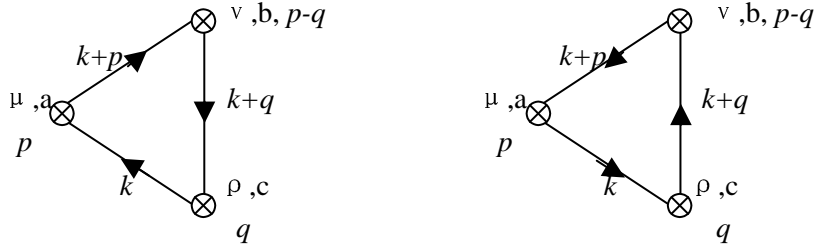,width=5in}}
 \centering
\begin{minipage}{5in}
   \caption{The triangle diagrams of constituent quarks.}
\end{minipage}
\end{figure}

\begin{eqnarray}\label{3.19}
S_3^{(\rm{normal})}&=&(\frac{3}{8}g^2-\frac{N_c}{16\pi^2})mg_{_A}^2
 \int d^4xTr_f\{\na_\mu U\na^\mu U^{\dag}(\td{\chi}U^{\dag}
 +U\td{\chi}^{\dag})\}\nonumber \\
&&+\int d^4x\int\frac{d^4p}{(2\pi)^4}e^{-ip\cdot x}
  \{-\frac{g_{_A}^2}{2}\alpha_2(p^2)p^\mu
  Tr_f[\bar{V}^\nu(p)[\Delta_\mu(x),\Delta_\nu(x)]]
   \nonumber \\
&&+\frac{N_c}{24\pi^2m}\alpha_3(p^2)(g_{\mu\nu}p^2-p_\mu p_\nu)
  Tr_f[\bar{V}^\mu(p)\bar{V}^\nu(x)
       (\xi\td{\chi}^{\dag}\xi+\xi^{\dag}\chi\xi^{\dag})]
\nonumber \\
&&-\frac{N_cm}{16\pi^2}\kappa g_{_A}\alpha_4(p^2)
  Tr_f[\bar{V}^\mu(p),\Delta_\mu(x)]
       (\xi^{\dag}\td{\chi}\xi^{\dag}-\xi\td{\chi}^{\dag}\xi)]
   \nonumber \\
&&-\frac{N_c}{16\pi^2m}\kappa g_{_A}\alpha_5(p^2)(g_{\mu\nu}p^2-p_\mu
  p_\nu)Tr_f[\bar{V}^\mu(p),\Delta^\nu(x)]
       (\xi^{\dag}\td{\chi}\xi^{\dag}-\xi\td{\chi}^{\dag}\xi)]\}
  +...,
\end{eqnarray}
where the form factor $\alpha_i(i=2,3,4,5)$ are defined as follows,
\begin{eqnarray}\label{3.20}
\alpha_2(p^2)&=&g^2-\frac{N_c}{2\pi^2}\int_0^1dx\int_0^1dy\cdot
      x(1-xy)[1+\frac{m^2}{m^2-f(p^2)}+\ln(1-\frac{f(p^2)}{m^2})],
     \nonumber \\
\alpha_3(p^2)&=&\int_0^1dx\frac{6x(1-x)m^2}{m^2-x(1-x)p^2},
  \nonumber\\
\alpha_4(p^2)&=&6\pi^2g^2-\int_0^1dx\int_0^1dy\{2x
  \ln(1-\frac{f(p^2)}{m^2})+\frac{xy(1-x)p^2}{m^2-f(p^2)}\}, \\
\alpha_5(p^2)&=&\int_0^1dx\int_0^1dy\frac{2x(1-x)m^2}{m^2-f(p^2)},
 \nonumber
\end{eqnarray}
with
\begin{eqnarray}\label{3.21}
f(p^2)=x(1-x)(1-y)p^2.
\end{eqnarray}

Moreover, the following three-point effective action with abnormal
parity is related to the purpose of this present paper,
\begin{eqnarray}\label{3.22}
S_3^{(\rm{ab})}=\frac{iN_c}{2\pi^2}g_{_A}\ep^{\mu\nu\alpha\beta}
 \int\frac{d^4q_1}{(2\pi)^4}\frac{d^4q_2}{(2\pi)^4}\alpha_6(q_1^2,q_2^2)
q_{1\mu}Tr_f\{\bar{V}_\nu(q_1)\bar{V}_\alpha(q_2)\Delta_\beta(q_1+q_2)\}
+...,
\end{eqnarray}
where
\begin{eqnarray}\label{3.23}
\alpha_6(q_1^2,q_2^2)=\int_0^1dx\cdot x\int_0^1dy
\{1-(x+y)(1-x)(1-y)\frac{q_1^2+q_2^2}{2m^2}\}^{-1}
\end{eqnarray}
It has pointed out that, the abnormal part of these quark models
including spin-1 meson usually do not pose the same symmetry as
the QCD anomalous Ward identities\cite{anomal1}, and the authors
of ref.\cite{anomal2} have provided a method to subtract the
extra terms. However, in this present paper we only focus our
attention on anomalous $\rho\rightarrow\gamma\pi$ and
$\omega\rightarrow\gamma\pi$ decays. Thus we have notation
$\Delta_\beta(k)\rightarrow k_\beta\pi(k)$ in eq.~(\ref{3.22}).
Using this notation we have
\begin{eqnarray}\label{pro}
&&\ep^{\mu\nu\alpha\beta}
Tr_f\{q_{1\mu}V_\nu(q_1)\Gamma_\alpha(q_2)\Delta_\beta(k)
 +q_{2\mu}V_\nu(q_1)\Gamma_\alpha(q_2)\Delta_\beta(k)\}\nonumber\\
\rightarrow&&\ep^{\mu\nu\alpha\beta}e
Tr_f\{-q_{2\mu}V_\nu(q_1){\cal Q}\gamma_\alpha(q_2)\Delta_\beta(k)
 +q_{1\mu}{\cal Q}\gamma_\nu(q_1)V_\alpha(q_2)\Delta_\beta(k)\}
\end{eqnarray}
be invariant under infinite transformation
\begin{eqnarray}\label{pro1}
&&{\cal O}_\mu^a\rightarrow {\cal O}_\mu^a+f^{abc}{\cal O}_\mu^b\ep^c,
\hspace{0.6in} {\cal O}_\mu=V_\mu,\;\Delta_\mu,\nonumber \\
&&{\cal O}_\mu^0\rightarrow {\cal O}_\mu^0,\hspace{1.3in}
\gamma_\mu\rightarrow\gamma_\mu+i\pa_\mu\ep^0,
\end{eqnarray}
where $\ep^c,(c=1,2,...8)$ and $\ep^0$ are infinite parameters.
Therefore, here no extra terms need to be subtracted.

The four-point effective action also includes both of normal parity part
and abnormal parity part. Due to soft-pion theorem, the normal parity part
relating to this paper reads
\begin{eqnarray}\label{3.24}
S_4^{({\rm normal})}&=&\frac{N_c}{192\pi^2}g_{_A}^4\int d^4x
  Tr_f\{\na_\mu U\na^\mu U^{\dag}\na_\nu U\na^\nu U^{\dag}\}\nonumber\\
&&+\frac{N_c}{8\pi^2m}g_{_A}^2\int d^4x\int\frac{d^4q}{(2\pi)^4}
  e^{-iq\cdot x}(g_{\mu\nu}q_\sigma-g_{\mu\sigma}q_\nu)
  {\Big (}\alpha_7(q^2)Tr_f\{\{\bar{V}^\mu(q),
  (\xi\td{\chi}^{\dag}\xi+\xi^{\dag}\chi\xi^{\dag})\}\Delta^{\nu}(x)
   \Delta^{\sigma}(x)\}\nonumber \\
  &&+\alpha_8(q^2)Tr_f\{\bar{V}^\mu(q)\Delta^\nu(x)
  (\xi\td{\chi}^{\dag}\xi+\xi^{\dag}\chi\xi^{\dag})\Delta^{\sigma}(x)\}
  {\Big )}+...,
\end{eqnarray}
where
\begin{eqnarray*}
\alpha_7(q^2)&=&\int_0^1dx_1\cdot x_1^2\int_0^1dx_2(1-x_2)
  \frac{3-2x_1^2x_2(1+2x_1)(1-x_2)q^2/m^2}{[1-x_1^2x_2(1-x_2)q^2/m^2]^2},
   \nonumber \\
\alpha_8(q^2)&=&\int_0^1dx_1\cdot x_1^2\int_0^1dx_2(1-x_2)
  \frac{4(1-x_1)[3-4x_1^2x_2(1-x_2)q^2/m^2]}
   {[1-x_1^2x_2(1-x_2)q^2/m^2]^2}.
\end{eqnarray*}

For abnormal parity part, we only focus on photon to three-pseudoscalar
interaction in this paper. Thus using soft-pion theorem, we have
\begin{eqnarray}\label{3.25}
S_4^{\gamma\rightarrow 3\Phi}=-\frac{N_c}{48\pi^2}eg_{_A}^3\int
 \frac{d^4q_2}{(2\pi)^4}\cdots\frac{d^4q_4}{(2\pi)^4}
i\ep^{\mu\nu\alpha\beta}q_{2\nu}q_{3\alpha}q_{4\beta}A_\mu(q_2+q_3+q_4)
 Tr_f\{{\cal Q}\Phi(q_2)\Phi(q_3)\Phi(q_4)\},
\end{eqnarray}
where $\Phi$ is pseudoscalar meson fields, $A_\mu$ is photon field, ${\cal
Q}={\rm diag}\{2/3,-1/3,-1/3\}$ is charge operator of quark fields.

So far, all meson fields in above effective actions are still
non-physical. The physical pseudoscalar meson fields can be obtained via
field rescaling $P\rightarrow 2F_\pi^{-1}P$ (here $P$ denotes
pseudoscalar mesons, and we ignore the difference between $F_\pi$ and
$F_K$ since we focus on $\rho$ and $\omega$ physics only). For defining
physical vector meson fields, let us pay attention to kinetic term of
vector mesons, which can be obtained from
eq.~(\ref{3.14})
\begin{eqnarray}\label{3.26}
{\cal L}_{\rm kin}^{(V)}=-\frac{1}{8}g^2
 Tr_f\{(\pa_\mu V_\nu-\pa_\nu V_\mu)(\pa^\mu V^\nu-\pa^\nu V^\mu)\}
 +\frac{1}{4}g^2m_{_V}^2Tr_f\{V_\mu V^\mu\}.
\end{eqnarray}
Here
\begin{eqnarray}\label{3.27}
m_{_V}^2=\frac{m_0^2}{g^2}+\frac{N_c}{\pi^2g^2}m_{_V}^2
 \int_0^1dx\cdot x(1-x)\ln(1-\frac{x(1-x)m_{_V}^2}{m^2})
\end{eqnarray}
is a ``common'' mass parameter of vector mesons. It implies that
the masses of light vector meson octet are degenerated at chiral
limit and large $N_c$ limit. From eq.~(\ref{3.26}) we can see
that the physical vector meson fields can be obtained via field
rescaling $V_\mu\rightarrow g^{-1}V_\mu$.

\subsection{Vector meson dominant and KSRF sum rules}

The direct coupling between photon and vector meson resonances is
also yielded by the effects of quark loops. Therefore, when vector
meson resonances are treated as bound states of constituent quarks,
vector meson dominant will be yielded naturally instead of input.
At the leading order of large $N_c$ expansion the VMD vertex reads from
eq.~(\ref{3.14})
\begin{eqnarray}\label{3.28}
 {\cal L}_{\rm VMD}=-\frac{1}{2}e\int\frac{d^4q}{(2\pi)^4}e^{-iq\cdot x}
  f_{\rho\gamma}^{(0)}(q^2)(g_{\mu\nu}q^2-q_\mu q_\nu)
  A^\mu(x)Tr_f\{QV^\nu(q)\}.
\end{eqnarray}
where $f_{\rho\gamma}^{(0)}(q^2)\equiv g^{-1}\alpha_1(q^2)$. At the
leading order of momentum expansion, i.e.,
$f_{\rho\gamma}^{(0)}(q^2=0)$, the above equation is just
the expression of VMD proposed by Sakurai\cite{Current}.
In addition, at the leading order of large $N_c$ expansion the
$V-\Phi\Phi$ vertex (where $V$ denote vector mesons and $\Phi$ denotes
pseudoscalar mesons) reads from eqs.~(\ref{3.14}) and (\ref{3.19})
\begin{eqnarray}\label{3.29}
{\cal L}_{\rm V\Phi\Phi}=-\frac{i}{4}\int\frac{d^4q}{(2\pi)^4}
  e^{-iq\cdot x}f_{\rho\pi\pi}^{(0)}(q^2)(g_{\mu\nu}q^2-q_\mu q_\nu)
  Tr\{V^\mu(q)[\Phi(x),\pa^\nu\Phi(x)]\},
\end{eqnarray}
where
\begin{eqnarray}\label{3.30}
f_{\rho\pi\pi}^{(0)}(q^2)=\frac{1}{gF_\pi^2}
  [\alpha_1(q^2)-g_{_A}^2\alpha_2(q^2)]
\end{eqnarray}

It is well known that the KSRF(I) sum rule\cite{KSRF}
\begin{equation}\label{3.31}
 f_{\rho\gamma}(q^2)=f_{\rho\pi\pi}(q^2)F_\pi^2
\end{equation}
is expected to be available at the leading order of momentum expansion.
At this order we have
\begin{eqnarray}\label{3.32}
f_{\rho\gamma}^{(0)}(q^2=0)&=&g, \nonumber\\
f_{\rho\pi\pi}^{(0)}(q^2=0)&=&\frac{1}{gF_\pi^2}\{g^2+
 (\frac{N_c}{3\pi^2}-g^2)\}.
\end{eqnarray}
We can see that KSRF(I) sum rule is satisfied exactly for
$g=\sqrt{\frac{N_c}{3}}\frac{1}{\pi}$. Therefore,
$g=\sqrt{\frac{N_c}{3}}\frac{1}{\pi}$ (especially, $g\equiv\pi^{-1}$
for $N_c=3$) is favorite choice for the universal constant of the model.

\subsection{Low energy limit}

It is well known that, at very low energy, ChPT is a rigorous consequence
of the symmetry pattern of QCD and its spontaneous breaking. So that the
low energy limit of those model including meson resonances must
match with ChPT. The low energy limit of EMG model can be obtained via
integrating over vector meson resonances. It means that, at very low
energy, the dynamics of vector mesons are replaced by pseudoscalar meson
fields. Since there are no interaction of vector mesons at $O(p^2)$,
at very low energy, the equation of motion
$\delta{\cal L}/\delta V_\mu=0$ yields classics solution for vector mesons
are follow
\begin{equation}\label{3.33}
V_\mu=\frac{1}{m_{_V}^2}\times O(p^3){\rm terms},
\end{equation}
where $p$ is momentum of pseudoscalar at very low energy. Therefore, in
effective action $S_n$, the terms involving vector meson resonances are
$O(p^6)$ at very low energy and do not contribute to $O(p^4)$ low energy
coupling constants, $L_i(i=1,2,...,10)$. The low energy coupling constants
$L_i(i\neq 7,8)$ yielded by EMG model can be directly
obtained from effective actions in subsection 3.2,
\begin{eqnarray}\label{3.34}
L_1&=&\frac{1}{2}L_2=\frac{N_c}{384\pi^2}, \hspace{0.8in}
L_3=-\frac{N_c}{64\pi^2}+\frac{N_c}{192\pi^2}g_A^4, \nonumber \\
L_4&=&L_6=0, \hspace{1.3in}L_5=\frac{N_cm}{32\pi^2B_0}g_A^2\\
L_8&=&\frac{F_0^2}{128B_0m}(3-\kappa^2)+\frac{3m}{64\pi^2B_0}
  (\frac{m}{B_0}-\kappa g_A-\frac{g_A^2}{2}-\frac{B_0}{6m}g_A^2)
  +\frac{L_5}{2}.\nonumber \\
L_9&=&\frac{N_c}{48\pi^2}, \hspace{1.35in}
L_{10}=-\frac{N_c}{48\pi^2}+\frac{N_c}{96\pi^2}g_A^2. \nonumber
\end{eqnarray}
In fact, the above expression on $L_i$ have been obtained in some
previous refs.\cite{MG84,Esp90,Bijnens93} (besides of $L_8$).

The constants $L_7$ has been known to get dominant contribution
from $\eta_0$\cite{GL85a} and this contribution is suppressed by
$1/N_c$. If we ignore the $\eta-\eta^{\prime}$ mixing, we have
\begin{equation}\label{3.35}
  L_7=-\frac{f_\pi^2}{128m_{\eta^{\prime}}^2}.
\end{equation}

Thus the six free parameters, $g$(it has been fitted by KSRF sum
rule), $g_{_A}$(it has been fitted by $n\rightarrow
pe^-\bar{\nu}_e$ decay), $B_0$, $\kappa$, $m$ and
$m_{\eta^{\pr}}$ determine all ten low energy coupling constants
of ChPT. It reflects the dynamics constrains between those low
energy coupling constants. Moreover, if we take $m_u+m_d\simeq
10$MeV, we can obtain $B_0=\frac{m_\pi^2}{m_u+m_d}\simeq 2$GeV.
Then experimental values of $L_8$ constrains constituent quark
mass $m\simeq 480$MeV (a more detail fit is in sect. {\bf IX}).
The numerical results for these low energy constants are in table
1.

\begin{table}[pb]
\centering
 \begin{tabular}{ccccccccccc}
&$L_1$&$L_2$&$L_3$&$L_4$&$L_5$&$L_6$&$L_7$&$L_8$&$L_9$&$L_{10}$
  \\ \hline
ChPT&$0.52\pm 0.23$&$0.72\pm 0.24$&$-2.7\pm 1.0$&$-0.3\pm 0.5$&$1.4\pm
  0.5$&$-0.2\pm 0.3$&$-0.26\pm 0.15$
  &$0.47\pm 0.18$&$6.9\pm 0.7$&$-5.2\pm 0.3$\\
{EMG}&0.79&1.58&-4.25&0&$1.26$&0&$(-0.4\pm 0.1)^{b)}$&
 $0.47^{a)}$&6.33&-4.55
   \end{tabular}
\begin{minipage}{6in}
\caption{\small $L_i$ in units of $10^{-3}$, ${\mu}=m_\rho$. The
   experimental data is from ref.[24]. a)input. b)contribution
   from gluon anomaly.}
\end{minipage}
\end{table}

\subsection{Chiral expansion at $m_\rho$-scale}

Let us calculate the on-shell decay $\rho\rightarrow\pi\pi$ and
$\rho^0\rightarrow e^+e^-$ under the leading order of momentum
expansion and under including all order information of vector
meson momentum expansion respectively. Numerically, using
$g_{_A}=0.75$, $g=\pi^{-1}$ and $m=480$MeV, we have
\begin{eqnarray}\label{3.36}
&&f_{\rho\gamma}^{(0)}(q^2=0)=\pi^{-1}=0.318,\hspace{0.8in}
f_{\rho\pi\pi}^{(0)}(q^2=0)=(0.93\times 10^{-5}){\rm MeV}^{-2},
 \nonumber\\
&&f_{\rho\gamma}^{(0)}(q^2=m_\rho^2)=0.439,\hspace{1.1in}
f_{\rho\pi\pi}^{(0)}(q^2=m_\rho^2)=(1.14\times 10^{-5}){\rm MeV}^{-2}.
\end{eqnarray}
Then the decay widths are $\Gamma(\rho\rightarrow\pi\pi)=125$MeV and
$\Gamma(\rho^0\rightarrow e^+e^-)=4.35$MeV at leading order of momentum
expansion, and $\Gamma(\rho\rightarrow\pi\pi)=188$MeV and
$\Gamma(\rho^0\rightarrow e^+e^-)=8.29$MeV when high order contributions
of momentum expansion are included. It implies that the high order
contributions of momentum expansion are very important at vector meson
energy scale. Furthermore, it indicates that the chiral expansion
converge slowly at vector meson energy scale. This is an important feature
of chiral theory including vector meson resonances.

However, neither the results at the leading order of momentum expansion
nor the results including high order contributions of vector meson
momentum expansion match with experimental data,
$\Gamma_{\rm exp}(\rho\rightarrow\pi\pi)=150$MeV and
$\Gamma_{\rm exp}(\rho^0\rightarrow e^+e^-)=(6.77\pm 0.32)$MeV. The reason
is that, so far, the contributions from pseudoscalar meson loops are still
not included. We have pointed out in introduction that these $N_c^{-1}$
suppression effects yield about $20\%$ contribution. Therefore, in the
following sections, we will calculate the effects of pseudoscalar meson
loops for various physical processes.

\section{One-loop correction to $\rho$ decays}
\setcounter{equation}{0}

In this section we will calculate pseudoscalar one-loop correction to
$\rho\rightarrow\pi\pi$ and $\rho^0\rightarrow e^+e^-$ decays and provide
a complete prediction on these reactions. Since $m_\pi^2\ll
m_{_K}^2<m_\rho^2$, we will treat pion as massless particle but
$m_{_K}\neq 0$ in our calculation. In addition, we will omit the
difference between $m_{\eta_8}$ and $m_{_K}$, and between $F_\pi$ and
$F_{_K}$, since these differences are doubly suppressed by light quark
mass expansion and $N_c^{-1}$ expansion.

\subsection{Correction to $\rho\rightarrow\pi\pi$ decay}

To the next to leading order of $N_c^{-1}$ expansion, there are three
kinds of loop diagrams of pseudoscalar mesons need to be calculated
(fig.2). Recalling $m_\pi^2=0$ but $m_{\eta_8}^2=m_{_K}^2\neq 0$, we can
see that only $K$ and $\eta_8$ mesons can yield non-zero contribution of
tadpole diagram in fig. 2-a), since in dimensional regularization,
$\int d^{D}kk^{2n}\equiv 0 (n\geq -1)$. Due to the same formula, we have
$$\int d^{D}k\frac{k^{2s}}{k^2+m_\varphi^2+i\epsilon}\propto 
m_\varphi^{2s} \times{\rm quadratic\;divergence},\hspace{1in} (s\geq 1).$$
Thus in this present paper we will ignore all contribution from quartic
divergence or higher order divergences. Furthermore, in the following
calculation we can find that the lowest order divergence is quadratic
divergence. In means that the on-shell condition for pseudoscalar mesons
and soft-pion theorem in tree level are still available even in
calculation on pseudoscalar meson loops. In addition, when we calculate
two-point diagram of pion, we must include all chain-like diagrams in fig.
2-c). Meanwhile, when we calculate two-point diagram of $K$ or $\eta_8$,
we only need to calculate one-loop diagram in fig. 2-b). The reason is
that the two-point diagram of pion generates a large imaginary part of
${\cal T}$-matrix for $\rho$ and $\omega$ physics, but $K$ or $\eta_8$ do
not. If we only calculate one-loop diagram in fig. 2-b), a large error bar
for phase factor will be yielded and this error will affect the study on
pion-pion phase shift in section 7.

\begin{figure}[hptb]
\centerline{\psfig{figure=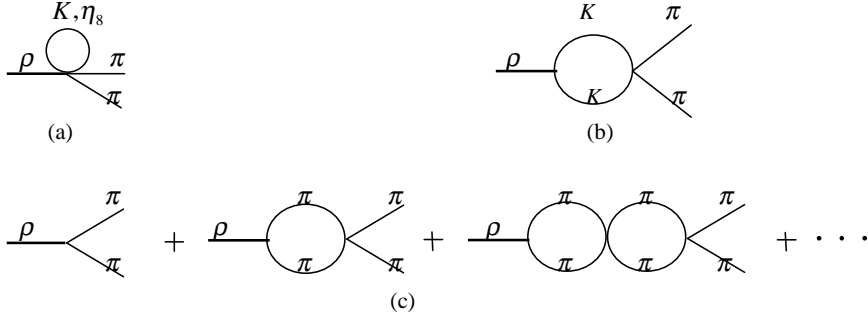,width=5in}}
\centering
\begin{minipage}{5in}
   \caption{One-loop diagrams for $\rho\rightarrow\pi\pi$ decay.
    a)Tadpole diagram of $K$ or $\eta_8$. b)Two-point diagram of
    $K$. c)Chain-like approximation of pion.}
\end{minipage}
\end{figure}

It is convenient to evaluate pseudoscalar meson loops in terms of
background field method. To expand pseudoscalar meson field around their
classic solution
\begin{eqnarray}\label{4.1}
U(x)=\bar{\xi}(x)e^{i\vphi}\bar{\xi}(x),
\hspace{1in}\bar{U}(\Phi)=\bar{\xi}^2(\Phi),
\end{eqnarray}
where background field $\bar{U}(\Phi)$ is solution of classic motion of
pseudoscalar mesons, $\delta{\cal L}/\delta U(x)=0$, $\vphi(x)$ is quantum
fluctuation fields around this classic solution.

Let us first calculate the tadpole diagram (fig. 2-a). Inserting
eq.~(\ref{4.1}) into effective action $S_2$ and $S_3$ and to retain terms
including two $\Phi$ and two quantum fields $\vphi$, the $\rho-4\Phi$
vertex reads,
\begin{eqnarray}\label{4.2}
{\cal L}_{\rho-4\Phi}=-\frac{1}{2F_\pi^2}\int\frac{d^4q}{(2\pi)^4}
  e^{-iq\cdot x}f_{\rho\pi\pi}^{(0)}(q^2)Tr_f\{\rho_{\mu\nu}(q)
  (\pa^\mu\Phi\vphi\pa^\nu\Phi\vphi+\vphi\pa^\mu\Phi\pa^\nu\Phi\vphi
  -\pa^\mu\Phi\pa^\nu\Phi\{\vphi,\vphi\})\}+\cdots.
\end{eqnarray}
where $q$ are four-momentum of $\rho$,
$\rho_{\mu\nu}(q)=q_\mu\rho_\nu(q)-q_\nu\rho_\mu(q)$, and $``\cdots''$
denotes those $\rho-4\Phi$ terms that only generate zero contribution,
e.g., $Tr_f\{\rho_{\mu\nu}(q)\pa^\mu\vphi\Phi\pa^\nu\vphi\Phi\}$.

In terms of the completeness relation of generators
$\lambda^a(a=1,2,\cdots,N^2-1)$ of $SU(N)$ group
\begin{eqnarray}\label{4.3}
&&Tr(\lambda^aA\lambda^aB)=-\frac{2}{N}Tr(AB)+2TrA TrB,
\nonumber \\
&&Tr(\lambda^aA)Tr(\lambda^aB)=2Tr(AB)-\frac{2}{N}TrA TrB,
\end{eqnarray}
we have
\begin{eqnarray}\label{4.4}
Tr_f\{\rho_{\mu\nu}(q)(\pa^\mu\Phi\lambda^a\pa^\nu\Phi\lambda^a
  +\lambda^a\pa^\mu\Phi\pa^\nu\Phi\lambda^a
  -\pa^\mu\Phi\pa^\nu\Phi\{\lambda^a,\lambda^a\})\}
  =-4NTr_f\{\rho_{\mu\nu}(q)\pa^\mu\Phi\pa^\nu\Phi\}.
\end{eqnarray}
From viewpoint of group theory, the contribution from $K$ and $\eta_8$ is
just one from pseudoscalar mesons in $SU(3)/SU(2)$ sector. Then tadpole
diagram correction to $\rho\rightarrow\Phi\Phi$ is
\begin{eqnarray}\label{4.5}
{\cal L}^{\rm (t)}_{\rho\Phi\Phi}&=&
  \frac{i}{4}\int\frac{d^4q}{(2\pi)^4}
  e^{-iq\cdot x}f_{\rho\pi\pi}^{(0)}(q^2)(g_{\mu\nu}q^2-q_\mu q_\nu)
  Tr_f\{\rho^\mu(q)[\Phi(x),\pa^\nu\Phi(x)]\}\int\frac{d^4k}{(2\pi)^4}
  \frac{4iF_\pi^{-2}}{k^2-m_{_K}^2+i\ep}\nonumber \\
&=&\frac{i}{4}\frac{\lambda m_{_K}^2}{4\pi^2F_\pi^2}
  \int\frac{d^4q}{(2\pi)^4}
  e^{-iq\cdot x}f_{\rho\pi\pi}^{(0)}(q^2)(g_{\mu\nu}q^2-q_\mu q_\nu)
  Tr_f\{\rho^\mu(q)[\Phi(x),\pa^\nu\Phi(x)]\},
\end{eqnarray}
where $\lambda$ is a constant absorbing quadratic divergence from
meson loops
\begin{eqnarray}\label{4.6}
\lambda=(\frac{4\pi\mu^2}{m^2})^{\ep/2}\Gamma(1-\frac{D}{2}).
\end{eqnarray}

In the following we will calculate the two-point diagram(TPD) of
pseudoscalar mesons. The leading order $\rho\Phi\Phi$ vertex of $N_c^{-1}$
expansion has been given in eq.~(\ref{3.29}). Here we need to further
obtain explicit 4-$\Phi$ vertex from the effective actions of section 3.
Inserting eq.~(\ref{4.1}) into $S_2$, $S_3^{\rm (normal)}$ and $S_4^{\rm
(normal)}$ and retaining terms with two $\Phi$ and two $\vphi$ fields, we
have
\begin{eqnarray}\label{4.7}
{\cal L}_{4\Phi}&=&\frac{1}{2}\int\frac{d^4q}{(2\pi)^4}
  e^{-iq\cdot x}[g_{\mu\nu}+\frac{N_c}{4F_\pi^2}
  (g^2-\frac{g_{_A}^4}{3\pi^2})
  (g_{\mu\nu}q^2-q_\mu q_\nu)]Tr_f\{\td{\Gamma}^\mu(q)
 [\vphi(x),\pa^\nu\vphi(x)]\} \nonumber \\
&&+\cdots,
\end{eqnarray}
with
\begin{eqnarray}\label{4.8}
\td{\Gamma}_\mu=\frac{1}{2F_\pi^2}[\Phi,\pa_\mu\Phi].
\end{eqnarray}
Here $``\cdots''$ denotes that those 4-$\Phi$ terms only yield zero
contribution due to
transveral operator $g_{\mu\nu}q^2-q_\mu q_\nu$ in $V-\Phi\Phi$ vertex,
such as $Tr_f\{[\pa_\mu\Phi,\vphi][\pa^\mu\Phi,\vphi]\}$. Then two-point
diagram correction to $\rho\Phi\Phi$ vertex is
\begin{eqnarray}\label{4.9}
{\cal L}_{\rho\Phi\Phi}^{\rm (TPD)}&=&\frac{1}{16F_\pi^2}
  \int\ frac{d^4q}{(2\pi)^4}
  f_{\rho\pi\pi}^{(0)}(q^2)(g_{\mu\nu}q^2-q_\mu q_\nu)
  [g_{\alpha\beta}+\frac{3}{4F_\pi^2}(g^2-\frac{g_{_A}^4}{3\pi^2})
   (g_{\alpha\beta}q^2-q_\alpha q_\beta)]
Tr_f\{\rho^\mu(q)[\lambda^a,\lambda^b]\}\nonumber \\&&\times
  Tr_f\{[\Phi(x),\pa^\alpha\Phi(x)][\lambda^a,\lambda^b]\}
  \int\frac{d^4k}{(2\pi)^4}\frac{ik^\nu}{k^2-m_{\vphi}^2+i\ep}
  \frac{i(2k+q)^{\beta}}{(k+q)^2-m_{\vphi}^2+i\ep}.
\end{eqnarray}
Using completeness relation of generators of $SU(N)$ group~(\ref{4.3}), we
have
\begin{eqnarray}\label{4.10}
Tr\{A[\lambda^a,\lambda^b]\}Tr\{B[\lambda^a,\lambda^b]\}
=-8NTr(AB)+8TrATrB.
\end{eqnarray}
Then taking $N=2$, we obtain $SU(2)$ (pion meson) correction as follow
\begin{eqnarray}\label{4.11}
{\cal L}_{\rm V\Phi\Phi}^{\rm TPD}[SU(2)]
  =\frac{i}{4}\int\frac{d^4q}{(2\pi)^4}e^{-iq\cdot x}
  f_{\rho\pi\pi}^{(0)}(q^2)\Sigma_\pi(q^2)(g_{\mu\nu}q^2-q_\mu q_\nu)
  Tr\{V^\mu(q)[\Phi(x),\pa^\nu\Phi(x)]\},
\end{eqnarray}
where
\begin{eqnarray}\label{4.12}
\Sigma_\pi(q^2)&=&\frac{q^2}{4\pi^2F_\pi^2}[1+\frac{3q^2}{4F_\pi^2}
  (g^2-\frac{g_{_A}^4}{3\pi^2})]\{\frac{\lambda}{6}
  +\int_0^1dx\cdot x(1-x)\ln{\frac{x(1-x)q^2}{m_{_K}^2}}\nonumber\\
  &&+\frac{i}{6}Arg(-1)\theta(q^2-4m_\pi^2)
 (1-\frac{4m_\pi^2}{q^2})^{3/2}\},
\end{eqnarray}
with
\begin{eqnarray}\label{4.13}
Arg(-1)&=&(1+2k)\pi,\hspace{0.7in}k=0,\pm 1,\pm 2,\cdots, \nonumber\\
\theta(x-y)&=&\left\{1,\hspace{1.1in}x>y; \atop
0,\hspace{1.1in}x\leq y.\right.
\end{eqnarray}
In eq.~(\ref{4.12}), the phase factor $(1-4m_\pi^2/q^2)^{3/2}$ denotes the
pion mass correction, which is yielded by unitarity of $S$-matrix(see
section 5). In addition, taking $N=3-2$ in eq.~(\ref{4.10}), we can obtain
$SU(3)/SU(2)$ ($K$-meson) correction
\begin{eqnarray}\label{4.14}
{\cal L}_{\rm V\Phi\Phi}^{\rm TPD}[SU(3)/SU(2)]
  =-\frac{i}{4}\int\frac{d^4q}{(2\pi)^4}e^{-iq\cdot x}
  f_{\rho\pi\pi}^{(0)}(q^2)\Sigma_K(q^2)(g_{\mu\nu}q^2-q_\mu q_\nu)
  Tr\{V^\mu(q)[\Phi(x),\pa^\nu\Phi(x)]\},
\end{eqnarray}
where
\begin{eqnarray}\label{4.15}
\Sigma_K(q^2)=\frac{m_{_K}^2}{8\pi^2F_\pi^2}[1+\frac{3q^2}{4F_\pi^2}
  (g^2-\frac{g_{_A}^4}{3\pi^2})]\{\lambda(1-\frac{q^2}{6m_{_K}^2})
  +\int_0^1dx(1-\frac{x(1-x)q^2}{m_{_K}^2})
   \ln(1-\frac{x(1-x)q^2}{m_{_K}^2})\}.
\end{eqnarray}

Comparing eq.~(\ref{4.11}) with eq.~(\ref{3.29}), we can see that every
two-point loop in chain-like approximation (fig. 2-c) contributes a factor
$-\Sigma_\pi(q^2)$. Therefore, to sum over all diagrams in fig. 2-c), we
obtain
\begin{eqnarray}\label{4.16}
{\cal L}_{\rm V\Phi\Phi}^{\pi-{\rm loop}}
  =-\frac{i}{4}\int\frac{d^4q}{(2\pi)^4}e^{-iq\cdot x}
  \frac{f_{\rho\pi\pi}^{(0)}(q^2)}{1+\Sigma_\pi(q^2)}
  (g_{\mu\nu}q^2-q_\mu q_\nu)Tr\{V^\mu(q)[\Phi(x),\pa^\nu\Phi(x)]\}.
\end{eqnarray}
Eq.~(\ref{4.16}) together with eqs.~(\ref{4.5}) and (\ref{4.13}) lead to
``complete'' $\rho-\pi\pi$ vertex to the next to leading order
of $N_c^{-1}$ expansion,
\begin{eqnarray}\label{4.17}
{\cal L}_{\rho\pi\pi}^{(c)}
  =-\frac{i}{4}\int\frac{d^4q}{(2\pi)^4}e^{-iq\cdot x}
   f_{\rho\pi\pi}^{(c)}(q^2)
  (g_{\mu\nu}q^2-q_\mu q_\nu)Tr\{\rho^\mu(q)[\pi(x),\pa^\nu\pi(x)]\},
\end{eqnarray}
where
\begin{eqnarray}\label{4.18}
f_{\rho\pi\pi}^{(c)}(q^2)=f_{\rho\pi\pi}^{(0)}(q^2)
  (\frac{1}{1+\Sigma_\pi(q^2)}+\Sigma_K(q^2)-
   \frac{\lambda m_{_K}^2}{4\pi^2F_\pi^2}).
\end{eqnarray}

\subsection{Correction to $\rho\rightarrow e^+e^-$ decay}

To the next to leading order of $N_c^{-1}$ expansion, we need consider
three kinds of loop diagrams of pseudoscalar mesons correction to VMD
vertex (fig.3).

\begin{figure}[hptb]
\centerline{\psfig{figure=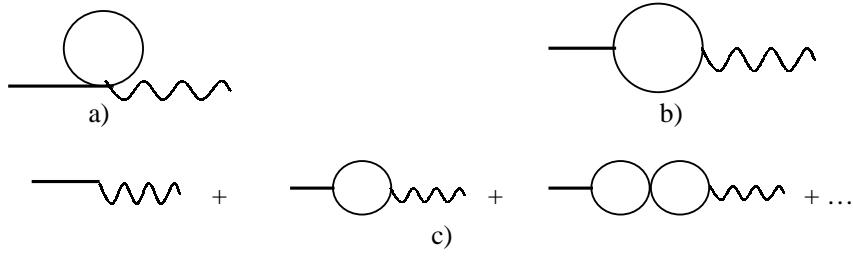,width=5in}}
\centering
\begin{minipage}{5in}
   \caption{One-loop diagrams for $\rho^0-\gamma$ mixing.
    a)Tadpole diagram of $K$ or $\eta_8$. b)Two-point diagram of
    $K$. c)Chain-like approximation of pion.}
\end{minipage}
\end{figure}

We first evaluate the tadpole contribution in fig.3-a). Inserting
eq.~(\ref{4.1}) into $S_2$, we obtain $\rho\gamma\vphi\vphi$ vertex as
follow
\begin{eqnarray}\label{4.19}
{\cal L}_{\rho\gamma\vphi\vphi}=\frac{1}{F_\pi^2}
  e\int\frac{d^4q}{(2\pi)^4}
  e^{-iq\cdot x}f_{\rho\gamma}^{(0)}(q^2)(g_{\mu\nu}q^2-q_\mu q_\nu)
  A^\mu(x)Tr_f\{\rho^\nu(q)(\frac{1}{2}{\cal Q}\{\vphi,\vphi\}
  -\vphi{\cal Q}\vphi)\}.
\end{eqnarray}
Then the tadpole correction of $K$ or $\eta_8$ mesons to VMD vertex is
\begin{eqnarray}\label{4.20}
{\cal L}_{\rm VMD}^{(t)}=\frac{1}{2}\frac{\lambda m_{_K}^2}
  {8\pi^2F_\pi^2}e\int\frac{d^4q}{(2\pi)^4}
  e^{-iq\cdot x}f_{\rho\gamma}^{(0)}(q^2)(g_{\mu\nu}q^2-q_\mu q_\nu)
  A^\mu(x)Tr_f\{{\cal Q}\rho^\nu(q)\}.
\end{eqnarray}

For calculating contribution of fig. 3-b) and fig. 3-c), the
$\gamma-\vphi\vphi$ vertex is necessary
\begin{eqnarray}\label{4.21}
{\cal L}_{\gamma\vphi\vphi}=-\frac{i}{2}e\int\frac{d^4q}{(2\pi)^4}
 e^{-iq\cdot x}[1+\frac{g}{2}
 f_{\rho\pi\pi}^{(0)}(q^2)(g_{\mu\nu}q^2-q_\mu q_\nu)]
 A^\mu(q)Tr_f\{{\cal Q}[\vphi(x),\pa^\nu\vphi(x)]\}.
\end{eqnarray}

Calculations on fig. 3-b) and fig. 3-c) are similar to one on
fig. 2. The final result on ``complete'' VMD vertex is
\begin{eqnarray}\label{4.22}
{\cal L}_{\rho\gamma}^{(c)}=-\frac{1}{2}e\int\frac{d^4q}{(2\pi)^4}
  e^{-iq\cdot x}f_{\rho\gamma}^{(c)}(q^2)(g_{\mu\nu}q^2-q_\mu q_\nu)
  A^\mu(x)Tr_f\{{\cal Q}\rho^\nu(q)\},
\end{eqnarray}
where
\begin{eqnarray}\label{4.23}
f_{\rho\gamma}^{(c)}(q^2)=f_{\rho\gamma}^{(0)}(q^2)
  (1-\frac{\lambda m_{_K}^2}{8\pi^2F_\pi^2})-f_{\rho\pi\pi}^{(0)}(q^2)
  [\frac{\Sigma_\pi^{\gamma}(q^2)}{1+\Sigma_\pi(q^2)}
   -\Sigma_K^{\gamma}(q^2)],
\end{eqnarray}
with
\begin{eqnarray}\label{4.24}
\Sigma_\pi^{\gamma}(q^2)&=&\frac{q^2}{4\pi^2}
  [1+\frac{g}{2}q^2f_{\rho\pi\pi}^{(0)}(q^2)]\{\frac{\lambda}{6}
  +\int_0^1dx\cdot x(1-x)\ln{\frac{x(1-x)q^2}{m_{_K}^2}} \nonumber\\
  &&+\frac{i}{6}Arg(-1)\theta(q^2-4m_\pi^2)
  (1-\frac{4m_\pi^2}{q^2})^{3/2}\},
  \nonumber \\
\Sigma_K^{\gamma}(q^2)&=&\frac{m_{_K}^2}{8\pi^2}
  [1+\frac{g}{2}q^2f_{\rho\pi\pi}^{(0)}(q^2)]
  \{\lambda(1-\frac{q^2}{6m_{_K}^2})
  +\int_0^1dx(1-\frac{x(1-x)q^2}{m_{_K}^2})
   \ln(1-\frac{x(1-x)q^2}{m_{_K}^2})\}.
\end{eqnarray}

\subsection{Cancellation of divergence}

From the above calculations we can find that only quadratic divergence
appears in one-loop contribution of pseudoscalar mesons. Since the present
model is a non-renormalizable effective theory, the divergences have to be
factorized, i.e., the parameter $\lambda$ has to be determined
phenomenologically.

The on-shell decay $\phi\rightarrow\pi\pi$ is forbidden by G
parity conservation and Zweig rule. Experiment also show that
branching ratios of this decay is very small,
$B(\phi\rightarrow\pi\pi)=(8\;{+5\atop -4})\times 10^{-5}$.
Theoretically, this decay can occur through photon-exchange or
$K$-loop(fig. 4). The latter two diagrams yield non-zero
imaginary part of decay amplitude. Thus the real part yielded by
the latter two diagrams must be very small. We can determine
$\lambda$ due to this requirement. From the calculation in the
above two subsection, we see that result yielded by the latter
two diagrams is proportional to a factor
\begin{equation}\label{4.25}
\lambda(\frac{p^2}{6}-m_{_K}^2)-\int_0^1dx\cdot [m_{_K}^2-x(1-x)p^2]
   \ln{(1-\frac{x(1-x)p^2}{m_{_K}^2})}.
\end{equation}
Then Zweig rule requires
\begin{equation}\label{4.26}
\{\lambda(\frac{p^2}{2}-m_{_K}^2)-Re\int_0^1dx\cdot
 [m_{_K}^2-x(1-x)p^2]\ln{(1-\frac{x(1-x)p^2}{m_{_K}^2})}\}
 |_{p^2=m_\phi^2}\simeq 0.
\end{equation}
Form the above equation, we obtain
\begin{equation}\label{4.27}
\lambda\simeq 0.54.
\end{equation}

\begin{figure}[hptb]
\centerline{\psfig{figure=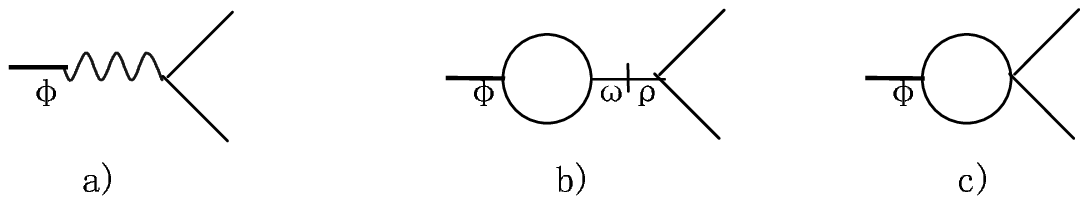,width=5in}}
\centering
\begin{minipage}{5in}
   \caption{Some diagrams for $\phi\rightarrow\pi\pi$ decay. The
   one-loop in figure b) and c) is $K$-loop.}
\end{minipage}
\end{figure}

\subsection{Numerical results}

The momentum-dependent widths for $\rho\rightarrow\pi\pi$ and
$\rho\rightarrow e^+e^-$ can be obtained from the ``complete''
vertices~(\ref{4.17}) and (\ref{4.22}),
\begin{eqnarray}\label{4.28}
\Gamma_{\rho\rightarrow\pi\pi}(q^2)&=&\frac{|f_{\rho\pi\pi}^{(c)}
  (q^2)|^2q^4}{48\pi}\sqrt{q^2}(1-\frac{4m_\pi^2}{q^2})^{3/2},
   \nonumber\\
\Gamma_{\rho\rightarrow e^+e^-}(q^2)&=&\frac{\pi\alpha_{\rm e.m.}^2}{3}
  |f_{\rho\gamma}^{(c)}(q^2)|^2\sqrt{q^2}.
\end{eqnarray}
Inputting $m=480$MeV, $g_{_A}=0.75$, $g=\pi^{-1}$, $\lambda=0.54$
and $\sqrt{q^2}=m_\rho=770$MeV, we obtain the on-shell decay
widths
\begin{eqnarray}\label{4.29}
&&\Gamma(\rho\rightarrow\pi\pi)=146.4{\rm MeV},\nonumber\\
&&\Gamma(\rho\rightarrow e^+e^-)=6.98{\rm KeV}.
\end{eqnarray}
These results agree with experimental data, $150$MeV and $(6.77\pm
0.32)$KeV, excellently. It also implies that the one-loop diagrams of
pseudoscalar mesons yield significant contribution at vector meson energy
scale.

\subsection{Loop correction to $\gamma\rightarrow\pi\pi$ vertex}

Here we will further calculate loop correction of pseudoscalar mesons to
$\gamma\rightarrow\pi\pi$ vertex, which is needed by studies in the
following section.

The calculation of loop correction to $\gamma\rightarrow\pi\pi$ vertex is
similar to one to $\rho\rightarrow\pi\pi$ vertex (fig. 5). The
$\gamma\Phi\Phi$ vertex and 4-$\Phi$ vertex has given in eqs.~(\ref{4.21})
and (\ref{4.7}) respectively. Here $\gamma$ to 4-$\Phi$ vertex is further
needed for evaluating the tadpole contribution in fig. 2-a),
\begin{eqnarray}\label{4.30}
{\cal L}_{\gamma-4\Phi}&=&-\frac{i}{2F_\pi^2}eA_\mu Tr_f\{[[{\cal
Q},\Phi],\vphi][\pa^\mu\Phi,\vphi]\}
-\frac{i}{F_\pi^2}e\int\frac{d^4q}{(2\pi)^4}e^{-iq\cdot x}
 gf_{\rho\pi\pi}^{(0)}(q^2)(q^2g_{\mu\nu}-q_\mu q_\nu)A^\mu(q)
  \nonumber\\&&\times
Tr_f\{{\cal Q}(\Phi\vphi\pa^\nu\Phi\vphi+\vphi\Phi\pa^\nu\Phi\vphi
  +\vphi\Phi\vphi\pa^\nu\Phi
  -\frac{3}{2}\Phi\pa^\nu\Phi\{\vphi,\vphi\}\}.
\end{eqnarray}

Then due to conservation of electromagnetic current, the ``complete''
$\gamma\rightarrow\pi\pi$ is
\begin{eqnarray}\label{4.31}
{\cal L}_{\gamma\pi\pi}^{(c)}=\frac{1}{2}
 \int\frac{d^4q}{(2\pi)^4}e^{-iq\cdot x}
 e\bar{F}_{\gamma\pi\pi}(q^2)A_\mu(q)Tr_f\{{\cal Q}[\pi,\pa^\mu\pi]\},
\end{eqnarray}
where $\bar{F}_{\gamma\pi\pi}(q^2)$ is nonresonant background
part of pion form factor,
\begin{eqnarray}\label{4.32}
\bar{F}_{\gamma\pi\pi}(q^2)&=&\frac{1}{1+\Sigma_\pi(q^2)}
 +\Sigma_K(q^2)-\frac{\lambda m_{_K}^2}{8\pi^2 F_\pi^2}
 +\frac{g}{2}q^2f_{\rho\pi\pi}^{(0)}(q^2)(\frac{1}{1+\Sigma_\pi(q^2)}
 +\Sigma_K(q^2)-\frac{3\lambda m_{_K}^2}{8\pi^2F_\pi^2})\nonumber\\
&&-\frac{3q^2}{4F_\pi^2}(g^2-\frac{g_{_A}^4}{3})
  [1+\frac{3q^2}{4F_\pi^2}(g^2-\frac{g_{_A}^4}{3})]^{-1}
  (\frac{\Sigma_\pi(q^2)}{1+\Sigma_\pi(q^2)}-\Sigma_k(q^2)).
\end{eqnarray}
Since $\Sigma_K(q^2)-\lambda m_{_K}^2/(8\pi^2 F_\pi^2)$ is $O(q^2)$,
$\bar{F}_{\gamma\pi\pi}(q^2=0)=1$ is still satisfied. It means that this
loop correction does not break $U(1)_{\rm e.m.}$ gauge invariance.

\section{Unitarity and Large $N_c$ Expansion}
\setcounter{equation}{0}

The unitarity of $S$-matrix, or optical theorem
\begin{eqnarray}\label{5.1}
{\rm Im}{\cal T}_{\beta,\alpha}=\frac{1}{2}\sum_{\rm all\;\gamma}
  {\cal T}_{\gamma,\alpha}{\cal T}_{\gamma,\beta}^*,
\end{eqnarray}
has to be satisfied for any well-defined quantum field theory,
where the ${\cal T}_{\beta,\alpha}$ transition amplitude, and
$\gamma$ denotes all possible intermediate states on mass shells.
It is well-known that a low energy effective meson theory should
be a well-defined perturbative theory in $N_c^{-1}$
expansion\cite{tH74}. Therefore, we can expand ${\cal T}$-matrix
in powers of $N_c^{-1}$ expansion,
\begin{eqnarray}\label{5.2}
{\cal T}=\sum_{n=0}^{\infty}{\cal T}_n,\hspace{1in}
{\cal T}_n\sim O((\sqrt{N_c})^{-n}).
\end{eqnarray}
Then the unitarity of $S$-matrix for a low energy effective meson theory
can be proved order by order in powers of $N_c^{-1}$ expansion,
\begin{eqnarray}\label{5.3}
{\rm Im}({\cal T}_{\beta,\alpha})_n=\frac{1}{2}\sum_{\rm all\;\gamma}
 \sum_{m\leq n}({\cal T}_{\gamma,\alpha})_m
 ({\cal T}_{\gamma,\beta}^*)_{n-m}.
\end{eqnarray}
In this present paper, we will discuss the unitarity of
$S$-matrix of several chiral quark model at vector meson energy
scale. In particular, as an example, we will directly examine
unitarity in EMG model via forward scattering of $\rho$-meson.

From effective action $S_2$, we can see that $F_\pi$ and $g$ are both
$O(\sqrt{N_c})$. So that every meson external line is $O(1/\sqrt{N_c})$
and every meson internal line is $O(1/N_c)$. In addition, from effective
action $S_n$ we also see that every vertex is $O(N_c)$. Therefore, any
transition amplitudes with $n_{_V}$ vertex, $n_e$ external meson lines,
$n_i$ internal meson lines and $n_l$ loops of mesons are order
\begin{eqnarray}\label{5.4}
N_c^{1-n_l-n_e/2},
\end{eqnarray}
where relation $n_l=n_i-n_{_V}+1$ has been used.

Using the above power counting rule, we can obtain a well-known fact that
unitarity of $S$-matrix requires that on-shell transition amplitude from a
vector meson to any meson states must be real at the leading order of
$N_c^{-1}$ expansion.

From eq.~(\ref{3.30}), we can see that, at the leading order
of $N_c^{-1}$ expansion, the on-shell $\rho\rightarrow\pi\pi$ transition
amplitude is proportional to $f_{\rho\pi\pi}^{(0)}(m_\rho^2)$. Thus the
above theorem requires $\alpha_1(m_\rho^2)$ and $\alpha_2(m_\rho^2)$
defined in eqs.~(\ref{3.15}) and (\ref{3.20}) are real. In the other
words, it means that, {\bf in chiral quark model including vector meson
resonances, the unitarity of $S$-matrix requires constituent quark mass}
${\bf m>m_\rho/2}$. In fact, this point is insured in EMG model by
$m\simeq 480$MeV which is fitted by low energy limit of the model.
However, in many other version of chiral quark
model\cite{Esp90,Chan85,ENJL,AE99}, the constituent quark mass are
constrained around 300MeV. Obviously, this value does not yield unitarity
$S$-matrix for single-vector meson processes, since all form factor
defined in section 3, $\alpha_i(q^2)$, are not real for $q^2\sim
m_\rho^2$. It also means that, in those model, the chiral expansion at
vector meson energy scale is not convergent.

In the following we will explicitly examine the eq.~(\ref{5.1}) in
forward scattering of $\rho-$meson up to three-loop level of mesons. The
examination on other processes can be performed similarly.
For $\alpha=\beta=\rho$ in eq.~(\ref{5.1}), $<\gamma|=<\pi\pi|$ is
dominant. The for forward scattering of $\rho$-meson, eq.~(\ref{5.1})
becomes
\begin{eqnarray}\label{5.7}
\frac{2}{(2\pi)^4}{\rm Im}{\cal T}_{\rho,\rho}
 =\Gamma(\rho\rightarrow\pi\pi)=\frac{|f_{\rho\pi\pi}^{(c)}(m_\rho^2)|^2
 m_\rho^5}{48\pi}(1-\frac{4m_\pi^2}{m_\rho})^{3/2}.
\end{eqnarray}
The both of ${\cal T}_{\rho,\rho}$ and $f_{\rho\pi\pi}^{(c)}$ can be
expanded in series of $N_c^{-1}$ expansion,
\begin{eqnarray}\label{5.8}
{\cal T}_{\rho,\rho}=\sum_{n=0}^{\infty}({\cal T}_{\rho,\rho})_{2n}
\hspace{1in}
f_{\rho\pi\pi}^{(c)}=\sum_{n=0}^{\infty}(f_{\rho\pi\pi}^{(c)})_{2n+1}.
\end{eqnarray}

\begin{figure}[hptb]
\centerline{\psfig{figure=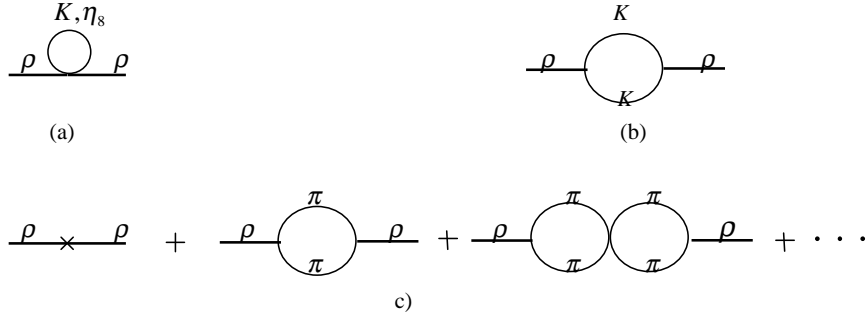,width=5in}}
\centering
\begin{minipage}{5in}
   \caption{One-loop diagrams correcting to $\rho$ propagator.
    a)Tadpole diagram of $K$ or $\eta_8$. b)Two-point diagram of
    $K$. c)Chain-like approximation of pion.}
\end{minipage}
\end{figure}

At the leading order, Im$({\cal T}_{\rho,\rho})_0=0$ is obviously
satisfied. For obtaining the imaginary part of $({\cal T}_{\rho,\rho})_2$,
we need to calculate meson loop correction in fig. 5. The calculations are
similar to one in section (I$\!$V.B), and the result is
\begin{eqnarray}\label{5.9}
{\cal L}_{\rho\rho}^{1-loop}=\frac{1}{2}\int\frac{d^4q}{(2\pi)^4}
 e^{-iq\cdot x}[f_{\rho\pi\pi}^{(0)}(q^2)]^2
 \{\frac{q^4\Sigma_\pi^{(0)}(q^2)}{1+\Sigma_\pi(q^2)}
   -q^2m_{_K}^2\Sigma_K^{(0)}(q^2)\}(g_{\mu\nu}q^2-q_\mu q_\nu)
  \rho_i^\mu(q)\rho_i^{\nu}(x),
\end{eqnarray}
where
\begin{eqnarray}\label{5.10}
\Sigma_\pi^{(0)}(q^2)&=&\frac{1}{8\pi^2}\{\frac{\lambda}{6}
  +\int_0^1dx\cdot x(1-x)\ln{\frac{x(1-x)q^2}{m_{_K}^2}}
  +\frac{i}{6}Arg(-1)\theta(q^2-4m_\pi^2)
  (1-\frac{4m_\pi^2}{q^2})^{3/2}\},
  \nonumber \\
\Sigma_K^{(0)}(q^2)&=&\frac{1}{16\pi^2}\{\lambda(1-\frac{q^2}
  {6m_{_K}^2})+\int_0^1dx(1-\frac{x(1-x)q^2}{m_{_K}^2})
   \ln(1-\frac{x(1-x)q^2}{m_{_K}^2})\}.
\end{eqnarray}
Then taking $Arg(-1)=-\pi$, we obtain
\begin{eqnarray}\label{5.11}
\frac{2}{(2\pi)^4}{\rm Im}({\cal T}_{\rho,\rho})_2
 =\frac{[f_{\rho\pi\pi}^{(0)}(m_\rho^2)]^2
 m_\rho^5}{48\pi}(1-\frac{4m_\pi^2}{m_\rho})^{3/2}.
\end{eqnarray}
Noting that $(f_{\rho\pi\pi}^{(c)})_1=f_{\rho\pi\pi}^{(0)}$, we can see
the eq.~(\ref{5.7}) is satisfied at the next to leading order.

\begin{figure}[hptb]
\centerline{\psfig{figure=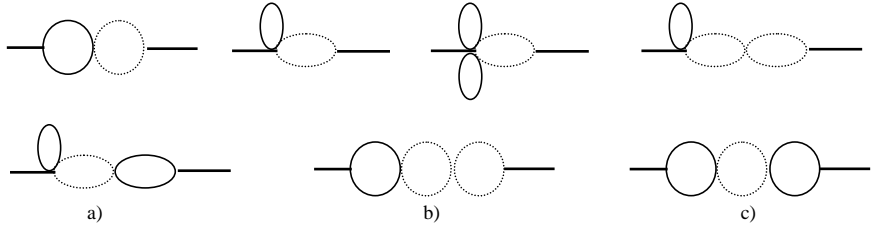,width=5in}}
\centering
\begin{minipage}{5in}
   \caption{Two-loop and three-loop diagrams correcting to $\rho$
   propagator. Here loops with solid line denote $K$ or $\eta_8$ loops,
   and loops with dash line denote pion loops. It should be pointed out
   that in fig. a), b) and c), the positions of pion loop and $K$ loop
   can exchange.}
\end{minipage}
\end{figure}

If the two-loop diagrams and three-loop diagrams in fig. 6 are
further considered, we can prove the eq.~(\ref{5.7}) is satisfied up to
$O(N_c^{-3})$. The details of calculation are omitted here.

Finally, eq.~(\ref{5.7}) also tell us that the ``complete''
propagator of $\rho$-meson is
\begin{eqnarray}\label{5.12}
\Delta_{\mu\nu}^{(\rho)}(q^2)=\frac{-i}{q^2-\td{m}_{\rho}^2
 +i\sqrt{q^2}\Gamma_{\rho}(q^2)}(g_{\mu\nu}-
 (\propto\frac{q_\mu q_\nu}{m_{\rho}^2})\;{\rm term}),
\end{eqnarray}
where $\Gamma_{\rho}(q^2)=\Gamma_{\rho\rightarrow\pi\pi}(q^2)$ given in
eq.~(\ref{4.28}). Since vector meson is treated at tree level, the
$(\propto q_\mu q_\nu/m_{\rho}^2)$ term in propagator is unimportant. Then
propagator~(\ref{5.12}) is just well-known Breit-Wigner formula for
spin-1 resonances.

Because the width in $\rho$-resonance (possessing a complex
pole) propagator~(\ref{5.12}) is momentum-dependent, it must be addressed
that {\bf the mass parameter $\td{m}_\rho$ is not the physical mass
$m_\rho=770$MeV.} Let us interpret this point briefly. Empirically,
the physical mass of resonance is defined as position of pole(real value)
in relevant scattering cross section, or theoretically, it should be
defined as real part of complex pole possessed by resonance. It is
well-known that the width of $\rho$-resonance is generated by pion
loops. For a simple VMD model, the leading order of $\rho-\pi\pi$ coupling
is independent of $q^2$. Thus one has $\Gamma_\rho^{(VMD)}(q^2)\propto
\sqrt{q^2}$, and due to equation
\begin{eqnarray}\label{5.14}
q^2-\td{m}_\rho^2+i\frac{\Gamma_\rho}{m_\rho}q^2=0,
\end{eqnarray}
we obtain the complex pole of $\rho$-resonance is
$q^2=m_\rho^2(1-i\ep+O(\ep^2))$ with $\ep=\Gamma_\rho/m_\rho\simeq 0.19$.
The result yields $\td{m}_\rho=m_\rho\sqrt{1+\ep^2}=784$MeV. In
particular, in the WCCWZ realization for vector meson fields used by this
paper, $\rho-\pi\pi$ coupling is proportional to $q^2$ at least. Hence one
has $\Gamma_\rho(q^2)\propto q^4\sqrt{q^2}$ at least, and complex pole
equation
\begin{eqnarray}\label{5.15}
q^2-\td{m}_\rho^2+i\frac{\Gamma_\rho}{m_\rho^5}q^6=0.
\end{eqnarray}
It yields $\td{m}_\rho=m_\rho\sqrt{1+3\ep^2}=810$MeV, which poses a
significant correction.

The above discussions imply that: 1) For resonance with large
width, the mass parameter in its propagator is different from its
physical mass. The correction is proportional to the ratio of
resonant width to physical mass. 2) The mass in the original
effective lagrangian only emerges as a parameter instead of a
physical quantity measured directly in experiment. 3) The choice
of mass parameter is relied on the choice of model. But the
physical quantity must be independent of this choice.

Since in our result all hadronic couplings include all order information
of the chiral perturbative expansion and one-loop effects of pseudoscalar
mesons, the momentum-dependence of $\Gamma_\rho(q^2)$ is very complicate.
It is difficult to determine $\td{m}_\rho$ via the above method. The
prediction on $\td{m}_\rho$ will be found in section {\bf V$\!$II}.

\section{$\rho^0-\omega$ Mixing and $\omega\rightarrow\pi^+\pi^-$ decay}
\setcounter{equation}{0}

The investigation of $\rho^0-\omega$ mixing, which is considered
as the important source of charge symmetry
breaking\cite{Miller90} in nuclear physics, has been an active
subject\cite{CB87}-$\!$\cite{WY00}. The mixing amplitude for
on-mass-shell of vector mesons has been observed directly in the
measurement of the pion form-factor in the time-like region from
the process $e^+e^-\rightarrow\pi^+\pi^-$\cite{OC98,Barkov85}.
For roughly twenty years, $\rho^0-\omega$ mixing amplitude was
assumed constant or momentum independent even if $\rho$ and
$\omega$ have the space-like momenta, far from the on-shell
point. Several years ago, this assumption was firstly questioned
by Goldman {\sl et. al.}\cite{GHT}, and the mixing amplitude was
found to be significantly momentum dependent within a simple
quark loop model. Subsequently, various authors have argued such
momentum dependence of the $\rho^0-\omega$ mixing amplitude by
using various approaches\cite{Momenta,OC94}. In particular, the
authors of ref.\cite{OC94} has pointed out that $\rho^0-\omega$
mixing amplitude must vanish at $q^2=0$(where $q^2$ denotes the
four-momentum square of the vector mesons) within a broad class
of model. This point will be also examined in EMG model.

In the most recent references\cite{CB87,Connell97,Bernicha94}, the
$\omega\rightarrow\pi^+\pi^-$ decay was treated as being dominant
via $\rho$-resonance exchange, and the direct $\omega\pi^+\pi^-$
coupling is neglected. It has been pointed out in
ref.\cite{Maltman96} that the neglect of $\omega$ ``direct''
coupling to $\pi^+\pi^-$ is not valid. It can be naturally
understood since $\pi\pi$ can make up of vector-isovector system,
whose quantum numbers are same to $\rho$ meson. Thus in an
effective lagrangian based on chiral symmetry, every $\rho$ field
can be replaced by $\pi\pi$ and does not conflict with symmetry.
Although authors of ref.\cite{Maltman96} also pointed out that
the present experimental data still can not be used to separate
``direct'' $\omega\pi\pi$ coupling from $\omega-\rho$ mixing
contribution in model-independent way, it is very interesting to
perform a theoretical investigation on ``direct'' $\omega\pi\pi$
coupling contribution.

It has been known that $\omega\rightarrow\pi^+\pi^-$ decay amplitude
receive the contribution from two sources: isospin symmetry breaking due
to $u-d$ quark mass difference and electromagnetic interaction. In section
3 we have shown that VMD in meson physics is natural consequence of the
present formalism instead of input. The $\rho\rightarrow e^+e^-$ decays
are also predicted successfully. Therefore, the dynamics of
electromagnetic interactions of mesons has been well established, and the
calculation for $\omega\rightarrow\pi^+\pi^-$ decay from the transition
$\omega\rightarrow\gamma\rightarrow\rho\rightarrow\pi\pi$ and ``direct''
$\omega\rightarrow\gamma\rightarrow\pi\pi$ is straightforward. Then we can
determine isospin breaking parameter $\delta m_q\equiv m_d-m_u$ via
$\omega\rightarrow\pi^+\pi^-$ decay. This parameter is urgently wanted by
determination of light quark mass ratios.

\subsection{Relevant effective vertices at tree level}

Here we first abstract those vertices relating to
$\omega\rightarrow\pi\pi$ decay from effective action in section III.

The $\rho^0-\omega$ mixing vertex, which breaks isospin symmetry, is
included in eq.~(\ref{3.19}),
\begin{eqnarray}\label{6.1}
{\cal L}_{\omega\rho}^{(0)}=\frac{N_c}{6\pi^2g^2}\frac{m_u-m_d}{m}
  \int\frac{d^4q}{(2\pi)^4}e^{-iq\cdot x}\alpha_3(q^2)(g_{\mu\nu}q^2
  -q_\mu q_\nu)\omega^\mu(q)\rho^{0\nu}(x)+O((m_d-m_u)^2).
\end{eqnarray}

The isospin symmetry unbroken vertex-$\Phi\Phi$ vertex (including
$\rho\pi\pi$, $\Delta I=0$ $\rho KK$ and $\omega KK$ vertices) and
4-$\Phi$ vertex has been given in eqs.~(\ref{3.29}) and (\ref{4.7})
respectively. The isospin symmetry broken vertex-$\Phi\Phi$ vertex
(including $\omega\pi\pi$, $\Delta I=0$ $\rho KK$ and $\omega KK$
vertices) can be abstracted from $S_3^{\rm (normal)}$ and $S_4^{\rm
(normal)}$,
\begin{eqnarray}\label{6.2}
{\cal L}_{V\Phi\Phi}^{(\Delta I=1)}=\frac{iN_c}{6\pi^2F_\pi^2}
 \frac{m_u-m_d}{m}\int\frac{d^4q}{(2\pi)^4}e^{-iq\cdot x}s(q^2)
 (g_{\mu\nu}q^2-q_\mu q_\nu)Tr_f\{\lambda_3V^\mu(q)
  \Phi(x)\pa^\nu\Phi(x)\},
\end{eqnarray}
where $\lambda_3={\rm diag}\{1,-1,0\}$, and form factor $s(q^2)$ is
defined as
\begin{eqnarray}\label{6.3}
s(q^2)=g^{-1}\{\alpha_3(q^2)+\frac{3}{4}g_{_A}^2(\alpha_7(q^2)-
 \frac{\alpha_8(q^2)}{2})\}.
\end{eqnarray}

The isospin symmetry broken 4-pseudoscalar vertex is proportional to
$m_d-m_u$, which is included in the following lagrangian
\begin{eqnarray}\label{6.4}
{\cal L}_{4\vphi}^{(\Delta I=1)}=
\frac{F_\pi^2}{8}B_0Tr_f\{{\cal M}(U+U^{\dag})\}
+\frac{N_cm}{(4\pi)^2}g_A^2Tr_f\{\na_\mu U\na^\mu U^{\dag}
 ({\cal M}U^{\dag}+U{\cal M})\}.
\end{eqnarray}

It can be found that the all relevant vertices are free parameter $\kappa$
independent. Moreover, we can see that every vector$\rightarrow\vphi\vphi$
in eq.(~\ref{6.2}) includes an transversal operator
$(g_{\mu\nu}q^2-q_\mu q_\nu)$ (where $q_\mu$ denotes four-momenta of
vector mesons). Thus the first term of eq.(~\ref{6.4}) does not contribute
to $\omega\rightarrow\pi^+\pi^-$ decay via pseudoscalar meson loops. This
transversal operator also constrains that the vertices with one of factors
$K\bar{K}$, $\pa_\mu K\pa^\mu\bar{K}$ and $K\pa^2\bar{K}$ do not
contribute to $\omega\rightarrow\pi^+\pi^-$ decay via pseudoscalar meson
loops. Then isospin symmetry breaking $KK\pi\pi$ vertices can explicitly
read as,
\begin{eqnarray}\label{6.5}
{\cal L}_{KK\pi\pi}^{(\Delta I=1)}=\frac{16N_c}{\pi^2F_\pi^4}g_A^2m
(m_uK^+\pa_\mu K^--m_dK^0\pa_\mu\bar{K}^0)\pi^+\pa^\mu\pi^-.
\end{eqnarray}

\subsection{Loop correction to ``direct'' $\omega\pi\pi$ vertex}

The loop correction of pseudoscalar mesons to ``direct'' $\omega\pi\pi$
vertex is similar to one to $\rho\pi\pi$ vertex (fig. 2). The
contribution from tadpole diagram is obtained via integrating quantum
fields in $\omega-4\Phi$ vertex, which is abstracted from $S_3$ and $S_4$,
\begin{eqnarray}\label{6.6}
{\cal L}_{\omega\pi\pi}^{(tad)}&=&\frac{N_c}{6\pi^2F_\pi^4}
  \frac{m_u-m_d}{m}\int\frac{d^4q}{(2\pi)^4}e^{iq\cdot x}s(q^2)
 (g_{\mu\nu}q_\sigma-g_{\mu\sigma}q_\nu)\omega^\mu(q)
  \int\frac{d^4k}{(2\pi)^4}\frac{i}
  {k^2-m_{_K}^2+i\ep} \nonumber \\&&\times\sum_{a=4}^8
 {\Big (}Tr_f\{\lambda_3(\pa^\nu\pi\lambda^a[\lambda^a,\pa^\sigma\pi]
  +\lambda^a[\lambda^a,\pa^\nu\pi]\pa^\sigma\pi)\}\nonumber \\&&
  \;\;\;\;+\frac{1}{2}Tr_f\{I_2\pa^\nu\pi\pa^\sigma\pi
 (\lambda^a\lambda^a{\cal M}+{\cal
 M}\lambda^a\lambda^a+\lambda^a{\cal M}\lambda^a)\}{\Big )}\nonumber \\
&=&-\frac{5\lambda m_{_K}^2}{12\pi^2F_\pi^2}\frac{iN_c}{6\pi^2F_\pi^2}
  \frac{m_u-m_d}{m}\int\frac{d^4q}{(2\pi)^4}e^{-iq\cdot x}s(q^2)
 (q^2\delta_{\mu\nu}-q_\mu q_\nu)\omega^\mu(q)
  Tr_f\{\lambda_3\pi(x)\pa^\nu\pi(x)\},
\end{eqnarray}
where $I_2={\rm diag}\{1,1,0\}$.

The ``direct'' $\omega\pi\pi$ coupling effective action yielded by
$K$-loop (fig. 2-b) can be evaluated as
\begin{eqnarray}\label{6.7}
iS_{\omega\pi\pi}^{(K-loop)}&=&-\int d^4xd^4y
  <0|T\{{\cal L}_{\rho KK}^{(\Delta I=0)}(x)
  {\cal L}_{\pi\pi KK}^{(\Delta I=1)}(y)+
  {\cal L}_{\rho KK}^{(\Delta I=1)}(x)
  {\cal L}_{\pi\pi KK}^{(\Delta I=0)}(y)\}|0>.
\end{eqnarray}
The result is
\begin{eqnarray}\label{6.8}
{\cal L}_{\omega\pi\pi}^{(K-loop)}&=&-\frac{iN_c}{6\pi^2F_\pi^2}
  \frac{m_u-m_d}{m}\int\frac{d^4q}{(2\pi)^4}e^{-iq\cdot x}
 \{6m^2f_{\rho\pi\pi}^{(0)}(q^2)[1+\frac{3q^2}{4\pi^2f_\pi^2}
  (g^2-\frac{g_{_A}^4}{3\pi^2})]^{-1}-s(q^2)\}
  \Sigma_K(q^2)\nonumber \\&&\hspace{0.5in}\times
  (q^2\delta_{\mu\nu}-q_{\mu}q_{\nu})\omega^\mu(q)
  Tr_f\{\lambda_3\pi(x)\pa^\nu\pi(x)\}.
\end{eqnarray}
The calculation on chain-like approximation correction of pion to
$\omega\pi\pi$ coupling is still similar to one on $\rho\rightarrow\pi\pi$
decay,
\begin{eqnarray}\label{6.9}
{\cal L}_{\omega\pi\pi}^{\pi-loop}=\frac{iN_c}{6\pi^2F_\pi^2}
 \frac{m_u-m_d}{m}\int\frac{d^4q}{(2\pi)^4}e^{-iq\cdot x}
 \frac{s(q^2)}{1+\Sigma_\pi(q^2)}(g_{\mu\nu}q^2-q_\mu q_\nu)
  \omega^\mu(q)Tr_f\{\lambda_3\pi(x)\pa^\nu\pi(x)\}.
\end{eqnarray}

\subsection{Loop correction to $\omega-\rho^0$ mixing}

The loop correction of pseudoscalar mesons to $\omega-\rho^0$ mixing is
similar to one to $\rho$-propagator (fig. 5). The contribution from
tadpole diagram is obtained via integrating quantum pseudoscalar fields in
$\omega\rho^0\vphi\vphi$ vertex, which is abstracted from $S_3^{\rm
(normal)}$,
\begin{eqnarray}\label{6.10}
{\cal L}_{\omega\rho^0}^{(tad)}&=&
 -\frac{N_c}{12\pi^2g^2F_\pi^2m}\int\frac{d^4q}{(2\pi)^4}
  e^{-iq\cdot x}\alpha_3(q^2)(g_{\mu\nu}q^2-q_\mu q_\nu)\omega^\mu(q)
 \rho^{0\nu}(x) \nonumber \\&&\times\int\frac{d^4k}{(2\pi)^4}\frac{i}
  {k^2-m_{_K}^2+i\ep} \sum_{a=4}^{8}
  Tr_f\{\lambda^3(\lambda^a\lambda^a{\cal M}+{\cal M}\lambda^a\lambda^a
  +\lambda^a{\cal M}\lambda^a)\} \nonumber\\
&=&-\frac{\lambda m_{_K}^2}{6\pi^2F_\pi^2}\frac{N_c}{6\pi^2g^2}
  \frac{m_u-m_d}{m}\int\frac{d^4q}{(2\pi)^4}e^{-iq\cdot x}
  \alpha_3(q^2)(g_{\mu\nu}q^2-q_\mu q_\nu)\omega^\mu(q)\rho^{0\nu}(x).
\end{eqnarray}

The calculations on two-point diagram of $K$-meson (fig. 5-b) and
chain-like approximation of pion meson (fig. 5-c) are similar to one in
section {\bf V}. The results is
\begin{eqnarray}\label{6.11}
{\cal L}_{\omega\rho}^{(K-loop)}&=&\frac{N_c}{3\pi^2}\frac{m_u-m_d}{m}
  \int\frac{d^4q}{(2\pi)^4}e^{-iq\cdot x}f_{\rho\pi\pi}^{(0)}(q^2)
  s(q^2)\Sigma_K(q^2)q^2(g_{\mu\nu}q^2-q_\mu q_\nu)\omega^\mu(q)
 \rho^{0\nu}(x),\nonumber \\
{\cal L}_{\omega\rho}^{(\pi-loop)}&=&-\frac{N_c}{3\pi^2F_\pi^2}
 \frac{m_u-m_d}{m}\int\frac{d^4q}{(2\pi)^4}e^{-iq\cdot x}
 \frac{\Sigma_\pi^{(0)}(q^2)}{1+\Sigma_\pi(q^2)}
 f_{\rho\pi\pi}^{(0)}(q^2)s(q^2)q^4(g_{\mu\nu}q^2-q_\mu
 q_\nu)\omega^\mu(q)\rho^{0\nu}(x),
\end{eqnarray}
where $\Sigma_\pi^{(0)}(q^2)$ is defined in eq.~(\ref{5.10}).

\subsection{Electromagnetic correction}

Due to VMD, the electromagnetic interaction contributes to
$\omega\rightarrow\pi^+\pi^-$ decay through one-photon exchange.
The ``complete'' $\omega-\gamma$ mixing vertex can be obtained
via calculating loop contribution in fig. 4-a) and fig. 4-b). It
should be pointed out that, in this case, the pion loop
correction is very small due to suppression of isospin
conservation. Then we have
\begin{eqnarray}\label{6.12}
{\cal L}_{\omega\gamma}^{(c)}=-\frac{1}{6}e\int\frac{d^4q}{(2\pi)^4}
  e^{-iq\cdot x}f_{\omega\gamma}^{(c)}(q^2)(g_{\mu\nu}q^2-q_\mu q_\nu)
  A^\mu(x)Tr_f\{{\cal Q}V^\nu(q)\},
\end{eqnarray}
where
\begin{eqnarray}\label{6.13}
f_{\omega\gamma}^{(c)}(q^2)=f_{\rho\gamma}^{(0)}(q^2)
  (1-\frac{\lambda m_{_K}^2}{8\pi^2F_\pi^2})+f_{\rho\pi\pi}^{(0)}(q^2)
  \Sigma_K^{\gamma}(q^2),
\end{eqnarray}
where $\Sigma_\pi^{(\gamma)}(q^2)$ is defined in eq.~(\ref{4.24}).
Then eqs.~(\ref{4.22}) and (~\ref{6.12}) will lead to
$\omega-\rho^0$ mixing at order of $\alpha_{\rm e.m.}$ through
the transition process
$\omega\rightarrow\gamma\rightarrow\rho^0$, which is
\begin{eqnarray}\label{6.14}
{\cal L}_{\omega\rho}^{\rm e.m.}=\frac{1}{12}e^2\int
  \frac{d^4q}{(2\pi)^4}
  e^{-iq\cdot x}f_{\rho\gamma}^{(c)}(q^2)f_{\omega\gamma}^{(c)}(q^2)
  (g_{\mu\nu}q^2-q_\mu q_\nu)\omega^\mu(q)\rho^{0\nu}(x).
\end{eqnarray}
Moreover, eqs.~(\ref{4.31}) and (\ref{6.12}) also lead to
``direct'' $\omega\pi\pi$ coupling at the order of $\alpha_{\rm
e.m.}$ through the transition process
$\omega\rightarrow\gamma\rightarrow\pi^+\pi^-$, which is
\begin{eqnarray}\label{6.15}
{\cal L}_{\omega\pi\pi}^{\rm e.m.}=\frac{1}{12}e^2\int
 \frac{d^4q}{(2\pi)^4}e^{-iq\cdot x}\bar{F}_{\rho\pi\pi}(q^2)
 f_{\omega\gamma}^{(c)}(q^2)(g_{\mu\nu}-\frac{q_\mu
 q_\nu}{q^2})\omega^\mu(q)
 Tr_f\{\lambda_3\pi(x)\pa^\nu\pi(x)\}.
\end{eqnarray}

\subsection{$\rho^0-\omega$ mixing and $\omega\rightarrow\pi^+\pi^-$
decay}

Equation~(\ref{6.14}) together with eqs.~(\ref{6.10}) and (\ref{6.11})
give the ``complete'' $\omega-rho^0$ mixing vertex as follow:
\begin{eqnarray}\label{6.16}
{\cal L}_{\omega\rho}^{(c)}=\int\frac{d^4q}{(2\pi)^4}
  e^{-iq\cdot x}\Theta_{\omega\rho}(q^2)(g_{\mu\nu}q^2-q_\mu q_\nu)
  \omega^\mu(q)\rho^{0\nu}(x),
\end{eqnarray}
where
\begin{eqnarray}\label{6.17}
\Theta_{\omega\rho}(q^2)&=&\frac{N_c}{6\pi^2g^2}\frac{m_u-m_d}{m}
  \{\alpha_3(q^2)(1-\frac{\lambda m_{_K}^2}{6\pi^2F_\pi^2})
   +2g^2q^2f_{\rho\pi\pi}^{(0)}(q^2)s(q^2)[\Sigma_K(q^2)
-\frac{q^2}{F_\pi^2}\frac{\Sigma_\pi^{(0)}(q^2)}{1+\Sigma_\pi(q^2)}]\}
  \nonumber\\&&+\frac{\alpha_{\rm e.m.}\pi}{3}
  f_{\rho\gamma}^{(c)}(q^2)f_{\omega\gamma}^{(c)}(q^2).
\end{eqnarray}

The complete ``direct'' $\omega\pi\pi$ vertex can be obtained via
summing eqs.(\ref{6.6}), (\ref{6.8}), (\ref{6.9}) and (\ref{6.15}),
\begin{eqnarray}\label{6.18}
{\cal L}_{\omega\pi\pi}^c&=&\frac{i}{2}
  \int\frac{d^4q}{(2\pi)^4}e^{-iq\cdot x}f_{\omega\pi\pi}(q^2)
  (q^2\delta_{\mu\nu}-q_\mu q_\nu)\omega^\mu(q)
  Tr_f\{\lambda_3\pi(x)\pa^\nu\pi(x)\}
\end{eqnarray}
where the form factor $f_{\omega\pi\pi}(q^2)$ is defined as follows:
\begin{eqnarray}\label{6.19}
f_{\omega\pi\pi}(q^2)&=&\frac{N_c}{3\pi^2F_\pi^2}\frac{m_u-m_d}{m}
  \{s(q^2)\left(\frac{1}{1+\Sigma_\pi(q^2)}
  -\frac{5\lambda m_{_K}^2}{12\pi^2F_\pi^2}\right)
       \nonumber \\&&
  -\left(6m^2f_{\rho\pi\pi}^{(0)}(q^2)[1+\frac{3q^2}{4\pi^2f_\pi^2}
  (g^2-\frac{g_{_A}^4}{3\pi^2})]^{-1}-s(q^2)\right)\Sigma_K(q^2)\}
 +\frac{2\alpha_{\rm e.m.}\pi}{3q^2}\bar{F}_{\gamma\pi\pi}(q^2)
  f_{\omega\gamma}(q^2).
\end{eqnarray}

Thus G-parity forbidden $\omega\rightarrow\pi^+\pi^-$ includes a
nonresonant background contribution, eq.(~\ref{6.18}), and $\rho$
resonance exchange contribution(eqs.(~\ref{4.17}) and (\ref{6.16})). The
decay width on $\omega$ mass-shell is
\begin{eqnarray}\label{6.20}
\Gamma(\omega\rightarrow\pi^+\pi^-)=\frac{m_\omega^5}{48\pi}
|\frac{m_\omega^2\Theta_{\omega\rho}(m_\omega^2)
 f_{\rho\pi\pi}^{(c)}(m_\omega^2)}
 {m_\omega^2-\td{m}_\rho^2+im_\omega\Gamma_\rho(m_\omega^2)}
  -f_{\omega\pi\pi}(m_\omega^2)|^2
 (1-\frac{4m_\pi^2}{m_\omega^2})^{3/2}.
\end{eqnarray}

For obtained the mass parameter of $\rho$-meson, $\td{m}_\rho$,
it is needed to study the difference between mass parameters of
$\rho^0$ and $\omega$. It is expected the masses of vector meson
octet is degenerated at chiral limit and large $N_c$ limit. In
general, the difference between mass parameters of $\rho^0$ and
$\omega$ is caused by three sources: $\rho^0-\omega$ mixing, VMD
and loop effects of pseudoscalar meson. The loop correction to
two-point vertex of $\rho$ has been obtained in eq.~(\ref{5.9}).
Similarly, since pion loop is suppressed by isospin conservation,
the loop correction to two-point vertex of $\omega$ is
\begin{eqnarray}\label{6.21}
{\cal L}_{\omega\omega}^{1-loop}=-\frac{1}{2}\int\frac{d^4q}{(2\pi)^4}
 e^{-iq\cdot x}[f_{\rho\pi\pi}^{(0)}(q^2)]^2m_{_K}^2\Sigma_K^{(0)}(q^2)
 q^2(g_{\mu\nu}q^2-q_\mu q_\nu)\omega^\mu(q)\omega^\nu(x).
\end{eqnarray}
Furthermore, due to eqs.~(\ref{4.22}), (\ref{6.12}) and (\ref{6.16}), the
calculation on VMD correction and $\rho^0-\omega$ mixing correction is
straightforward. In addition, since $\Gamma_\omega\ll m_\omega$, the mass
parameter of $\omega$ is almost just the physical mass of $\omega$. In the
other words, $\td{m}_\omega=m_\omega$ is a good approximation. Then the
final results are
\begin{eqnarray}\label{6.22}
m_\omega^2&=&m_{_V}^2+{\rm Re}\{\frac{q^4\Theta_{\omega\rho}^2(q^2)}
 {q^2-\td{m}_\rho^2+i\sqrt{q^2}\Gamma_\rho(q^2)}+
 \frac{\pi\alpha_{\rm e.m.}}{9}q^2[f_{\omega\gamma}^{(c)}(q^2)]^2
 -q^4[f_{\rho\pi\pi}^{(0)}(q^2)]^2m_{_K}^2
 \Sigma_K^{(0)}(q^2)\}|_{q^2=m_\omega^2},\nonumber\\
\td{m}_\rho^2&=&m_{_V}^2+{\rm Re}\{\frac{q^4\Theta_{\omega\rho}^2(q^2)}
 {q^2-m_\omega^2+i\sqrt{q^2}\Gamma_\omega}+
 \pi\alpha_{\rm e.m.}q^2[f_{\rho\gamma}^{(c)}(q^2)]^2\nonumber\\
&&+q^4[f_{\rho\pi\pi}^{(0)}(q^2)]^2\left(\frac{q^2\Sigma_\pi^{(0)}(q^2)}
  {1+\Sigma_\pi(q^2)}-m_{_K}^2\Sigma_K^{(0)}(q^2)\right)
 \}|_{q^2=m_\rho^2}.
\end{eqnarray}

Using the experimental data, $m_\omega=782$MeV and
$B(\omega\rightarrow\pi^+\pi^-)=(2.21\pm 0.30)\%$\cite{PDG98},
together with eq.(~\ref{6.20}), we obtain
\begin{eqnarray}\label{6.23}
 m_{_V}=782.5{\rm MeV},\hspace{1in}\td{m}_\rho=803{\rm MeV}
\end{eqnarray}
and
\begin{eqnarray}\label{6.24}
m_u-m_d=-(3.87\pm 0.21){\rm MeV}
\end{eqnarray}
at energy scale $\mu\sim m_\omega$. Here the error bar is from the
uncertainty in branch ratio of the process $\omega\rightarrow\pi^+\pi^-$.
In the standard way, the $\omega-\rho^0$ mixing amplitude is
\begin{eqnarray}\label{6.25}
\int\frac{d^4q}{(2\pi)^4}\Pi_{\omega\rho}(q^2)=
<\omega|\int d^4x{\cal L}_{\omega\rho}(x)|\rho>\;
\Longrightarrow\Pi_{\omega\rho}(q^2)&=&q^2\Theta_{\omega\rho}(q^2).
\end{eqnarray}
The off-shell $\omega-\rho^0$ mixing amplitude is obviously
momentum dependent, and vanished at $q^2=0$. This is consistent
with the argument by O'Connell {\sl et. al.} in ref.\cite{OC94}
that this mixing amplitude must vanish at the transition from
time-like to space-like four momentum within a broad class of
models. In addition, the value of isospin broken
parameter(~\ref{6.24}) leads on $\omega$ mass-shell
$\omega-\rho^0$ mixing amplitude as follow
\begin{equation}\label{6.26}
{\rm Re}\Pi_{\omega\rho}(m_\omega^2)=-(3888\pm 253){\rm MeV}^2,
 \hspace{0.5in}
{\rm Im}\Pi_{\omega\rho}(m_\omega^2)=-(2018\pm 122){\rm MeV}^2.
\end{equation}
In ref.\cite{OC98}, the on-shell mixing amplitude has extracted
from the $e^+e^-\rightarrow\pi^+\pi^-$ experimental data in a
model-dependent way. In eq.(~\ref{6.25}), the real part of
on-shell mixing amplitude agree with result of ref.\cite{OC98}.
The imaginary part, however, is much larger than one in
ref.\cite{OC98} which is around $-300$MeV$^2$. It must be pointed
out that, in ref.\cite{OC98} the author's analysis bases on a
model without ``direct'' $\omega\pi\pi$ coupling. Therefore, it is
insignificant to compare the value of on-shell mixing amplitude
of this the present paper with one of ref.\cite{OC98}.
Fortunately, the ratio between $\omega\rightarrow\pi\pi$ decay
amplitude and $\rho\rightarrow\pi\pi$ decay amplitude should be
model-independent. This value can test whether a model is right
or not. The on-shell mixing amplitude in ref.\cite{OC98} yields
\begin{equation}\label{6.27}
R_{\omega\rho}^{\rm exp}=\frac{<\pi^+\pi^-|\omega>}{<\pi^+\pi^-|\rho>}
       =-(0.0060\pm 0.0009)+(0.0322\pm 0.0050)i.
\end{equation}
The present paper predicts
\begin{equation}\label{6.28}
R_{\omega\rho}=\frac{m_\omega^2\Theta_{\omega\rho}(m_\omega^2)}
 {m_\omega^2-\td{m}_\rho^2+im_\omega\Gamma_\rho(m_\omega^2)}
  -\frac{f_{\omega\pi\pi}(m_\omega^2)}{f_{\rho\pi\pi}(m_\omega^2)}
  =-(0.0084\pm 0.0007)+(0.0323\pm 0.0021)i.
\end{equation}
We can see that this theoretical prediction agree with experimental
excellently.

If we take $f_{\omega\pi\pi}(q^2)=0$ in eq.(~\ref{6.20}), we have
$B(\omega\rightarrow\pi^+\pi^-)=(2.16\pm 0.19)\%$. So that the
contribution from interference between ``direct'' $\omega\pi\pi$
coupling and $\omega-\rho^0$ mixing is very small. The dominant
contribution is from $\rho$-resonance exchange. This result
indicates all previous studies which without ``direct''
$\omega\pi\pi$ coupling are good approximation even though the
neglecting the ``direct'' coupling is an $ad\;hoc$ assumption.

\section{$\omega\rightarrow\gamma\pi^0$ and
$\rho^\pm\rightarrow\gamma\pi^\pm$ decays}
\setcounter{equation}{0}

At the leading order of $N_c^{-1}$ expansion, the $\rho\gamma\pi$ and
$\omega\gamma\pi$ vertices are including in anomalous three-point
effective action~(\ref{3.22}). In momentum space, it is
\begin{eqnarray}\label{7.1}
\Pi_{V\gamma\pi}=\frac{N_c}{3\pi^2gF_\pi}eg_{_A}\alpha_6(q^2,0)
  \ep^{\mu\nu\alpha\beta}q_\alpha k_\beta\{\rho^i_\mu(q)A_\nu(k)
  \pi_i(q-k)+3\omega_\mu^i(q)A_\nu(k)\pi^0(q-k)\}.
\end{eqnarray}

The loop correction on these anomalous decays are similar to one on
$\rho\rightarrow\pi\pi$ decay. The contribution from tadpole diagram is
\begin{eqnarray}\label{7.2}
\Pi_{V\gamma\pi}^{tad}(q,k)&=&\frac{N_c}{4\pi^2 gF_\pi^3}g_{_A}
 \int\frac{d^4p}{(2\pi)^4}\frac{i\delta^{ab}}{p^2-m_{_K}^2+i\ep}
 \ep^{\mu\nu\alpha\beta}q_\alpha k_\beta\pi^{l}
  Tr_f{\Big \{}\frac{1}{3}(\rho_\mu^i(q)\lambda^i+3\omega_\mu(q)
   +eQA_\mu(q))\nonumber\\ &&\times
  (\rho_\nu^j(k)\lambda^j+3\omega_\nu(k)+eQA_\nu(k))
  ([\lambda^a, \lambda^l]\lambda^b+\lambda^a[\lambda^l,\lambda^b])
    \nonumber \\&&
  +e[(\rho_\mu^i(q)\lambda^i+3\omega_\mu(q)),
    ([\lambda^a,Q]\lambda^b+\lambda^a[Q,\lambda^b])]A_\nu(k)
  \lambda^l{\Big \}}\hspace{0.5in} (i,j,l=1,2,3;a,b=4,...,8)\nonumber\\
&=&-\frac{\lambda m_{_K}^2}{6\pi^2F_\pi^2}\Pi_{V\gamma\pi}(q,k).
\end{eqnarray}

Up to $O(p^4)$, the $\gamma$-3$\Phi$ vertex can be abstracted from
eqs.~(\ref{3.22}) and (\ref{3.25}),
\begin{eqnarray}\label{7.3}
\Pi_{\gamma\rightarrow 3\Phi}(k_1,k_2,k_3)=-\frac{iN_c}{2\pi^2F_\pi^3}
 eg_{_A}(1-\frac{g_{_A}^2}{3})\ep^{\mu\nu\alpha\beta}
 k_{1\nu}k_{2\alpha}k_{3\beta}A_\nu(k_1+k_2+k_3)
 Tr_f\{{\cal Q}\Phi(k_1)\Phi(k_2)\Phi(k_3)\}.
\end{eqnarray}
Then using eqs.~(\ref{3.29}) and (\ref{7.3}), the contribution
from two-point diagram of $K$-meson is
\begin{eqnarray}\label{7.4}
\Pi_{V\gamma\pi}^{K-loop}(q,k)&=&\frac{N_c}{8\pi^2F_\pi}eg_{_A}
  (1-\frac{g_{_A}^2}{3})f_{\rho\pi\pi}^{(0)}(q^2)\Sigma_K(q^2)
  q^2\ep^{\mu\nu\alpha\beta}q^2q_\alpha k_\beta\nonumber\\&&\times
  \{\rho^i_\mu(q)A_\nu(k)
  \pi_i(q-k)+3\omega_\mu^i(q)A_\nu(k)\pi^0(q-k)\}.
\end{eqnarray}
In addition, $\rho\gamma\pi$ vertex also receives contribution
from chain-like approximation of pion loops (the contribution to
$\omega\gamma\pi\pi$ is suppressed by isospin conservation). The
result is
\begin{eqnarray}\label{7.5}
\Pi_{\rho\gamma\pi}^{\pi-loop}(q,k)&=&-\frac{N_c}{8\pi^2F_\pi}eg_{_A}
  (1-\frac{g_{_A}^2}{3})f_{\rho\pi\pi}^{(0)}(q^2)
  \frac{\Sigma_\pi(q^2)}{1+\Sigma_\pi(q^2)}
  q^2\ep^{\mu\nu\alpha\beta}q^2q_\alpha k_\beta
  \rho^i_\mu(q)A_\nu(k)\pi_i(q-k).
\end{eqnarray}

Then eq.~(\ref{7.1}) together with eqs.~(\ref{7.2}), (\ref{7.4}) and
(\ref{7.5}) lead to ``complete'' $\rho\gamma\pi$ and $\omega\gamma\pi$
coupling to the next to leading order of $N_c^{-1}$ expansion at least,
\begin{eqnarray}\label{7.6}
\Pi_{V\gamma\pi}^{c}(q,k)&=&e\ep^{\mu\nu\alpha\beta}q_\alpha
  k_\beta\{f_{\rho\gamma\pi}(q^2)\rho_\mu^i(q)A_\nu(k)\pi_i(q-k)
+g_{\omega\gamma\pi}(q^2)\omega_\mu(q)A_\nu(k)\pi^0(q-k)\},\nonumber \\
f_{\rho\gamma\pi}(q^2)&=&\frac{N_c}{3\pi^2gF_\pi}g_{_A}
 (1-\frac{\lambda m_{_K}^2}{6\pi^2F_\pi^2})\alpha_6(q^2,0)
   \nonumber\\
&&-\frac{N_c}{8\pi^2F_\pi}g_{_A}(1-\frac{g_{_A}^2}{3})q^2
 f_{\rho\pi\pi}^{(0)}(q^2)\left(\frac{\Sigma_\pi(q^2)}
 {1+\Sigma_\pi(q^2)}-\Sigma_K(q^2)\right),\nonumber\\
f_{\omega\gamma\pi}(q^2)&=&\frac{N_c}{\pi^2gF_\pi}g_{_A}
 (1-\frac{\lambda m_{_K}^2}{6\pi^2F_\pi^2})\alpha_6(q^2,0)
+\frac{N_c}{8\pi^2F_\pi}g_{_A}(1-\frac{g_{_A}^2}{3})q^2
 f_{\rho\pi\pi}^{(0)}(q^2)\Sigma_K(q^2).
\end{eqnarray}
For $g_{_A}(\mu=m_\rho)=0.75$, the above results yield
\begin{eqnarray}\label{7.7}
B(\rho^\pm\rightarrow\pi^\pm\gamma)=4.75\times 10^{-4},
\hspace{1in}B(\omega\rightarrow\pi^0\gamma)=8.8\%.
\end{eqnarray}
These results agree with data\cite{PDG98}
$B(\rho^\pm\rightarrow\pi^\pm\gamma)=(4.5\pm 0.5)\times 10^{-4}$
and $B(\omega\rightarrow\pi^0\gamma)=(8.5\pm 0.5)\%$ very well.
Since in these anomalous decay, the effects of axial constant
$g_{_A}$ is the leading order, it can be checked whether
$g_{_A}=0.75$ or not, which is fitted by $\beta$ decay of
neutron. If we take $g_{_A}=1$ as a comparison, the theoretical
predictions are
\begin{eqnarray}\label{7.8}
B(\rho^\pm\rightarrow\pi^\pm\gamma)=8.51\times 10^{-4},
\hspace{1in}B(\omega\rightarrow\pi^0\gamma)=15.6\%.
\end{eqnarray}
These results obviously disagree with data. This point implies that the
EMG model are available not only on light flavour meson physics but also
on light flavour baryon physics.

\section{Pion form factor}
\setcounter{equation}{0}

The process $e^+e^-\rightarrow\pi^+\pi^-$ at energies lower than
the chiral symmetry spontaneously breaking scale contains very
important information on low energy hadron dynamics. It was an
active subject and studied continually during past fifty years.
Experimentally, the effects of the strong interaction in process
of $e^+e^-$ annihilation is obvious to provides a large
enhancement to production of pions in vector meson resonance
region\cite{Barkov85,OLYA,CMD,Augu73,HL81}. Theoretically,
however, the problem was not studied completely so far. In some
recent refrences\cite{Bena93,Bena97,Connell96}, the authors have
studied $e^+e^-\rightarrow\pi^+\pi^-$ cross section and
$l=1,\;I=1$ phase shift at vector meson resonance region by using
some simple phenomenological models. Each of them capture some
leading order effects of the chiral expansion and $N_c^{-1}$
expansion, and they are classified by different symmetry
realization for vector meson fields\cite{Brise97}. However, as we
shown in previous sections, it is not enough to consider the
leading order effects only at vector meson scale. Otherwise many
delicate features on vector meson physics will be lost.

Eqs.~(\ref{4.17}), (\ref{4.22}), (\ref{4.31}), (\ref{6.12}), (\ref{6.16})
and (\ref{6.18}) lead to the electromagnetic form factor of pion
as follows:
\begin{eqnarray}\label{8.1}
F_{\gamma\pi\pi}(q^2)=\bar{F}_{\gamma\pi\pi}(q^2)-
  \frac{q^4f_{\rho\gamma}^{(c)}(q^2)f_{\rho\pi\pi}^{(c)}(q^2)}
  {2(q^2-\td{m}_\rho^2+i\sqrt{q^2}\Gamma_\rho(q^2))}
  -\frac{q^4f_{\omega\gamma}^{(c)}(q^2)g_{\omega\pi\pi}^{(c)}(q^2)}
  {6(q^2-m_\omega^2+i\sqrt{q^2}\Gamma_\omega)},
\end{eqnarray}
where
\begin{eqnarray}\label{8.2}
g_{\omega\pi\pi}^{(c)}(q^2)=\frac{q^2\Theta_{\omega\rho}
 f_{\rho\pi\pi}^{(c)}(q^2)}{q^2-\td{m}_\rho^2+i\sqrt{q^2}
 \Gamma_\rho(q^2)}-f_{\omega\pi\pi}^{(c)}(q^2),
\end{eqnarray}
Here due to narrow width of $\omega$, we ignore the momentum-dependence of
$\Gamma_\omega$. In this form factor, we can see that the contributions of
resonance exchange accompany $q^4$ factor. Due to this reason, some
authors declared that the pion form factor in WCCWZ realization for vector
meson resonances exhibits an
unphysical high energy behaviour($\mu>m_\rho$). However, this conclusion
is wrong. It is caused by their wrong result for momentum-dependence of
$\Gamma_\rho(q^2)$ which is fitted by experimental instead of by dynamical
prediction. In fact, since $\sqrt{q^2}\Gamma_\rho(q^2)$ is proportional to
$q^6$ at least, we do not need to worry that the form factor has a bad
high energy behaviour. We can also see that there is a moment-dependent
non-resonant contribution. It together with the contribution of resonance
exchange determined the high energy behaviour of the factor. The
cross-section for $e^+e^-\rightarrow\pi^+\pi^-$ is given by(neglecting the
electron mass)
\begin{eqnarray}\label{8.3}
\sigma=\frac{\pi\alpha_{\rm e.m.}^2}{3}\frac{(q^2-4m_\pi^2)^{3/2}}
  {(q^2)^{5/2}}|F_\pi(q^2)|^2.
\end{eqnarray}

\begin{figure}[tph]
   \centerline{
   \psfig{figure=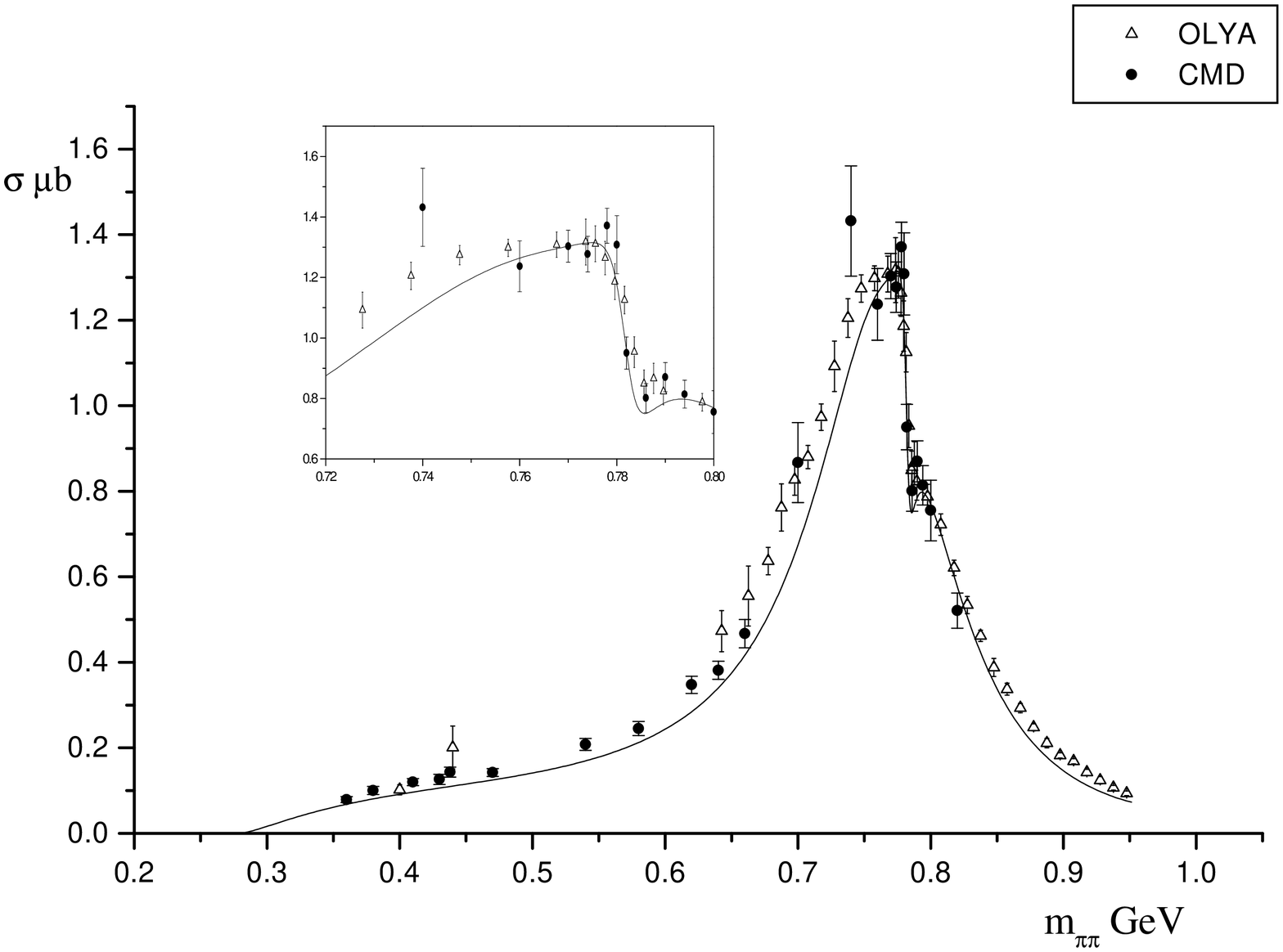,width=5.8in}}
\begin{center}
\begin{minipage}{5in}
   \caption{$e^+e^-\rightarrow\pi^+\pi^-$ cross section. The
  experimental data are from refs.[31,32].}
\end{minipage}
   \end{center}
\end{figure}

From definition of function $\alpha_i(q^2)$ in section {\bf III}
we can see this model is unitary only for $q^2<4m^2$. Thus the
effective prediction should be below $m_{\pi\pi}<2m=960$MeV. The
result is shown in fig. 7. We can see the prediction agree with
data well. Especially, the theoretical prediction in vector meson
energy region agree with data excellently. Although the mass
parameter $\td{m}_\rho=803$MeV in $\rho$ propagator is larger
than physical mass, the position of pole is localized in
$\sqrt{q^2}=772$MeV which is just the physical mass of $\rho$. It
strongly supports our above discussion and dynamical calculation.
It also implies that we must carefully distinguish the physical
mass difference of $\rho^0$ and $\omega$ from the difference of
mass parameter in effective lagrangian.

Let us give some further remarks on pion form factor~(\ref{8.2}).
In eq.~(\ref{8.2}), the form factors $\bar{F}_{\gamma\pi\pi}(q^2)$,
$f_{\rho\gamma}(q^2)$, etc., are all complex function instead of real
function. It is caused by one-loop effects of pions. Thus the
expression~(\ref{8.2}) can be rewritten as follows:
\begin{eqnarray}\label{8.4}
F_{\gamma\pi\pi}(q^2)=1+q^2b_1(q^2)e^{i\phi_1(q^2)}-
  \frac{q^4b_2(q^2)e^{i\phi_2(q^2)}}{2(q^2-\td{m}_\rho^2
  +i\sqrt{q^2}\Gamma_\rho(q^2))}-\frac{q^4b_3(q^2)e^{i\phi_3(q^2)}}
 {6(q^2-m_\omega^2+i\sqrt{q^2}\Gamma_\omega)}.
\end{eqnarray}
Here $b_i(q^2)(i=1,2,3)$ are three real function and
$\phi_i(q^2)(i=1,2,3)$ are three momentum-dependent phases. In
particular, $\phi_3(q^2=m_\omega^2)$, so called Orsay phase, has
been extracted from data as $100-125$
degrees\cite{Bena97,Connell96}. Our theoretical prediction is
$\phi_3(q^2=m_\omega^2)=116.5$ degrees. However, so far, the
phases $\phi_1(q^2)$ and $\phi_2(q^2)$ are not reported in any
literature. These momentum-dependent phases indicate that the
dynamics including loop effects of pseudoscalar mesons is
different from one only in tree level. In fig. 8, we given
theoretical curves of $\phi_i(q^2)$. They are indeed nontrivial.

\begin{figure}[hptb]
  \centerline{\psfig{figure=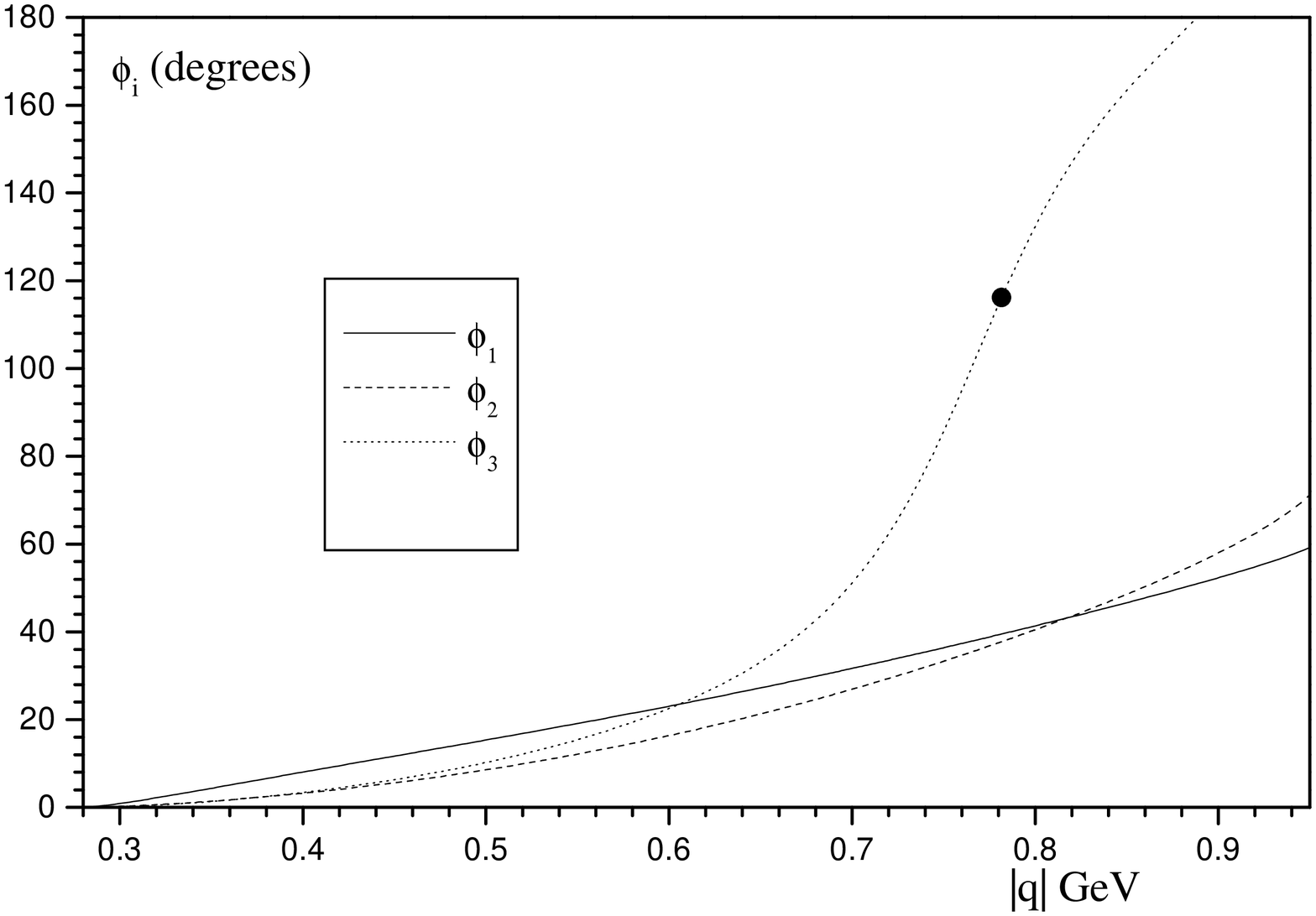,width=5.5in}}
\begin{center}
\begin{minipage}{5in}
   \caption{$\phi_i$ versus $m_{\pi\pi}$ in GeV. Here the solid line
denotes the phase shift of non-resonant background $\phi_1$, the dash line
denotes the phase shift $\phi_2$ in $\rho$ coupling and the dot line
denotes the phase shift $\phi_3$ in $\omega$ coupling. ``$\bullet$''
denotes Orsay phase.}
\end{minipage}
\end{center}

\end{figure}

Obviously, $F_\pi(q^2)$ is an analytic function in the complex $q^2$
plane, with a branch cut along the real axis beginning at the two-pion
threshold, $q^2=4m_\pi^2$. Time-reversal invariance and the unitarity of
the $S$-matrix requires that the phase of the form factor be that of
$l=1,\;I=1$ $\pi-\pi$ scattering\cite{GT58}. This last emerges as
$\pi-\pi$ scattering in the relevant channel is very nearly elastic from
threshold through $q^2\simeq (m_\pi+m_\omega)^2$\cite{HL81,HOP73}. In this
region of $q^2$, then, the form factor is related to the $l=1,\;I=1$
$\pi-\pi$ phase shift, $\delta_1^1$, via\cite{Gasi66}
\begin{eqnarray}\label{8.5}
F_{\gamma\pi\pi}(q^2)=e^{2i\delta_1^1(q^2)}F_{\gamma\pi\pi}^*(q^2),
\end{eqnarray}
so that
\begin{eqnarray}\label{8.6}
\tan{\delta_1^1(q^2)}=\frac{{\rm Im}
 F_{\gamma\pi\pi}(q^2)}{{\rm Re}F_{\gamma\pi\pi}(q^2)}.
\end{eqnarray}
The above is a special case of what is sometimes called the
Fermi-Watson-Aidzu phase theorem\cite{Gasi66,Watson55}. In fig. 9 and
fig. 10 we plot theoretical curves of the $l=1,\;I=1$ $\pi-\pi$ phase
shift
$\delta_1^1$ versus $m_{\pi\pi}$ and of $\sin{\delta_1^1}/p_\pi^3$ versus
$m_{\pi\pi}$ (where $p_\pi=\frac{1}{2}\sqrt{q^2-4m_\pi^2}$) respectively.
We omit the $\omega$ contribution from our plots of the phase of
$F_\pi(q^2)$ for comparing with time-like region pion form factor
data~\cite{Data73,Data74,Data77}. We have also assumed that $\delta_1^1$
is purely elastic in the regime shown, i.e., the loop effects of
$\omega-\pi$ are omitted. The curve predicts $\delta_1^1\rightarrow
90^\circ$ as $\sqrt{q^2}\rightarrow 774{\rm MeV}
\simeq m_\rho$, and $\delta_1^1>100^\circ$ for $\sqrt{q^2}>787$MeV. These
results agree with data very well.

\begin{figure}[hptb]
  \centerline{\psfig{figure=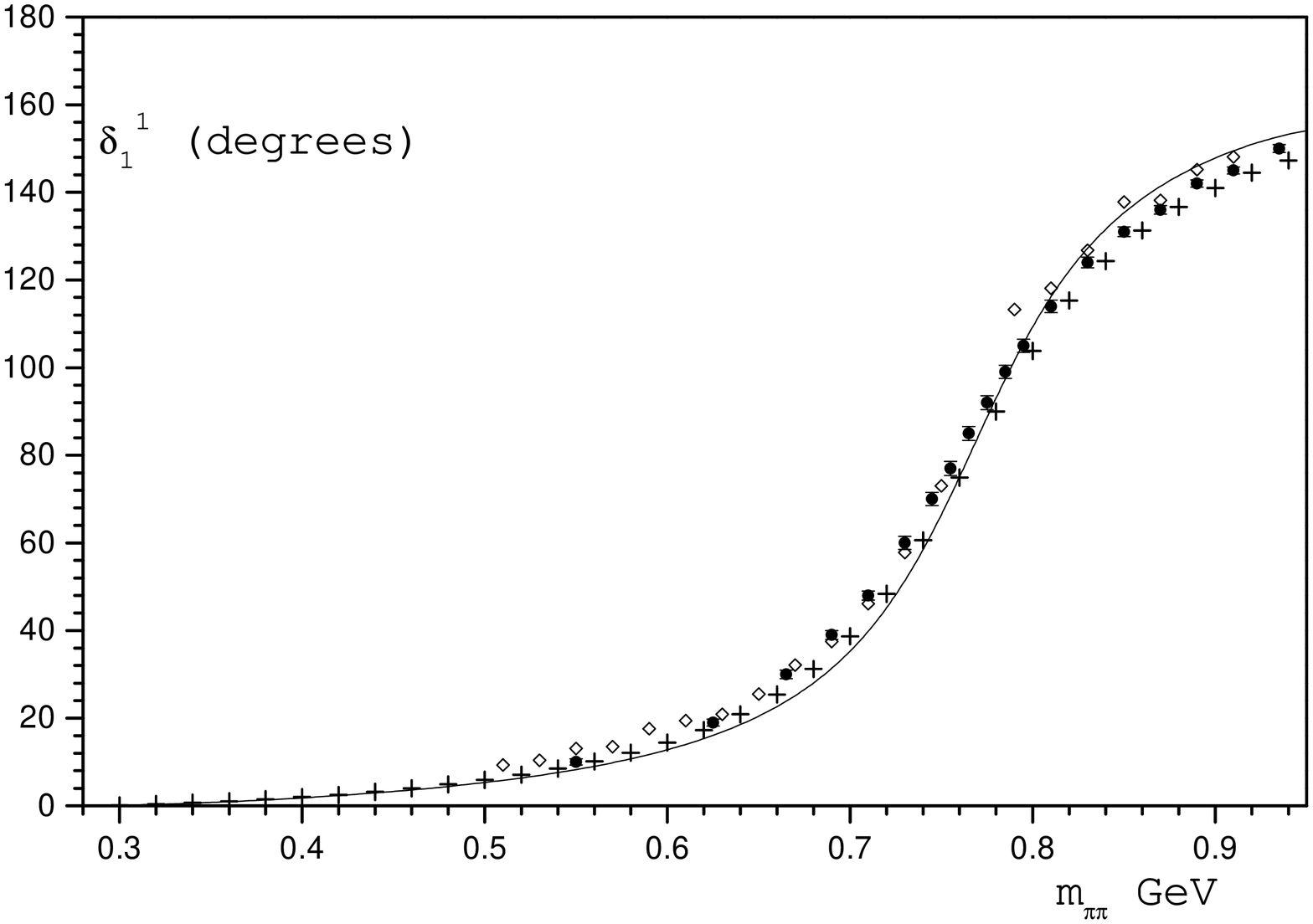,width=5.5in}}
\begin{center}
\begin{minipage}{5in}
\caption{The $l=1,\;I=1$ $\pi-\pi$ scattering phase shift. The solid
circle point are these from [44], the hollow diamond point are from
[45] and ``+'' denotes the points from [46]. Note that the $\omega$
contribution to the time-like pion form factor phase has been omitted, to
facilitate comparison with the empirical phase shift.}
\end{minipage}
   \end{center}
\end{figure}

Finally we discuss the near threshold behaviour of the form
factor. 1) The chiral perturbative theory predicts the form
factor at threshold to be $[F_{\gamma\pi\pi}(4m_\pi^2)]_{\rm
ChPT}=1.17\pm 0.01$, and ours, $[F_\pi(4m_\pi^2)]=1.145$, is
close to the ChPT result. 2) The electromagnetic radius of
charged pion has been determined to be $\sqrt{<r^2>_\pi}=0.657\pm
0.027$fm\cite{Dally82}, whereas the theoretical prediction in
this present paper is $\sqrt{<r^2>_\pi}=0.645$ fm. 3) The
Froggatt-Petersen phase shift function $\sin{\delta_1^1}/p_\pi^3$
is connected with the vector-isovector $\pi-\pi$ scattering
length $a_1^1$ through
\begin{eqnarray}\label{8.7}
a_1^1=\lim_{q^2\rightarrow 4m_\pi^2}\frac{\sin{\delta_1^1}}{p_\pi^3}.
\end{eqnarray}
Our theoretical prediction is $a_1^1=0.037$ in unit of $m_\pi^{-3}$. This
value is very close to experimental results from $K_{e4}$
data\cite{Nagel79,Rosselet77} using a Roy equation fit ($a_1^1=0.038\pm
0.002$) and ChPT prediction $a_1^1=0.037\pm 0.01$\cite{Knecht95} at the
two loop order (at $O(p^4)$).

To provide a brief summary on this section. In section {\bf V},
it has been revealed that the mass parameter in resonant
propagators should be different from its physical mass due to
large momentum-dependent width. In this section, this point is
confirmed by both of dynamical calculation and phenomenological
fit. It also tells us how to understand the mass splitting
between $\rho^0$ and $\omega$: Although the dynamical
calculations show that the mass parameter of $\rho$ in effective
lagrangian is even larger than one of $\omega$, the position of
pole localized in real axis give their right physical mass
splitting. The contribution to this mass splitting from
$\rho^0-\omega$ mixing is very small, the dominant contribution
is from one-loops effects of pseudoscalar mesons. The effective
field theory mechanics on $\rho^0-\omega$ mass splitting revealed
in this present paper is rather subtle. And it is another evident
to confirm again that the EMG model is sound.

\begin{figure}[hptb]
  \centerline{\psfig{figure=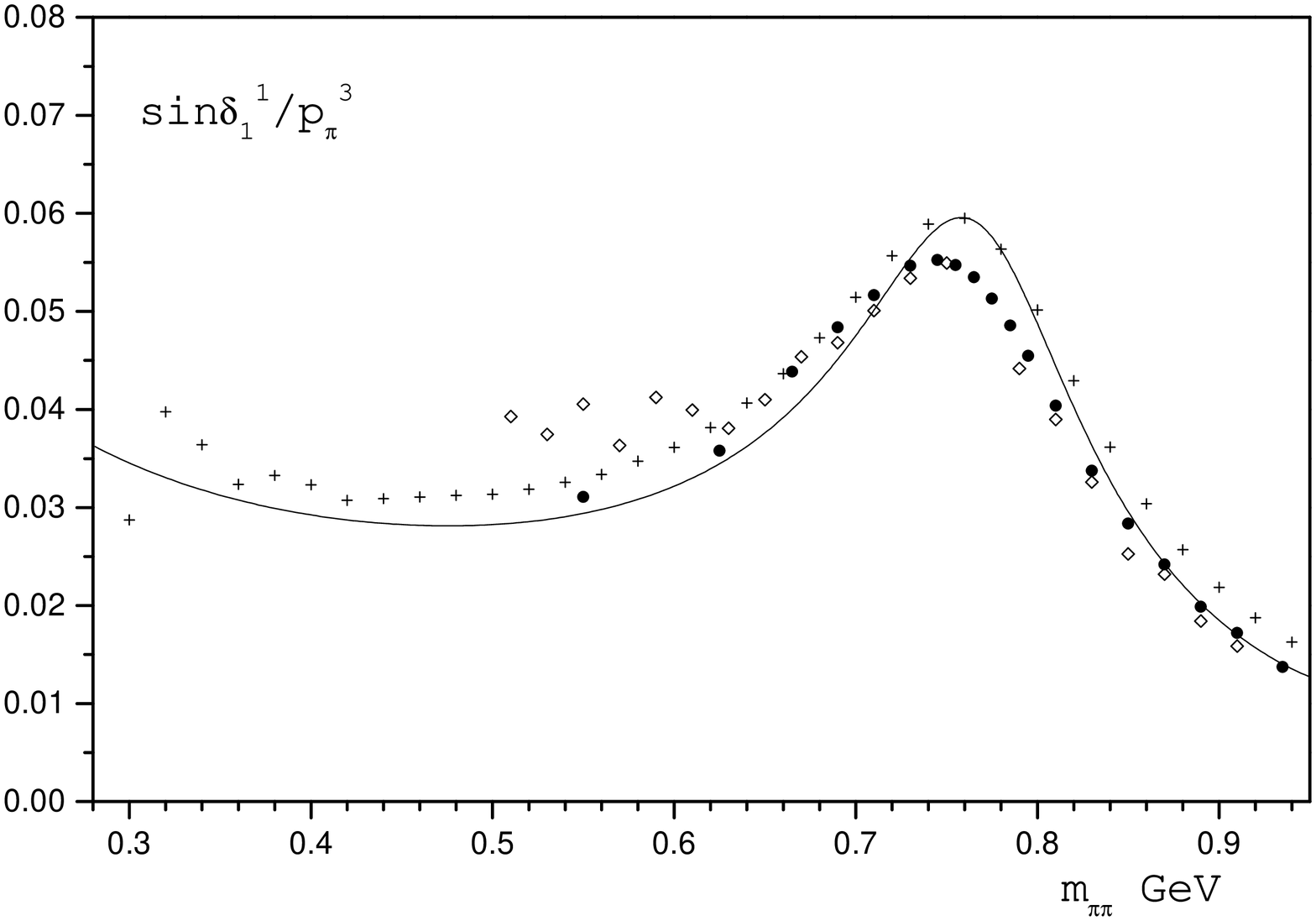,width=5.5in}}
\begin{center}
\begin{minipage}{5in}
\caption{Function $\sin{\delta_1^1}/p_\pi^3$ deduced from $\pi^+\pi^-$
phase shift; the function is given in units of $m_\pi^{-3}$. The solid
circle point are these from [44], the hollow diamond point are from
[45] and ``+'' denotes the points from [46].}
\end{minipage}
   \end{center}
\end{figure}

\section{Light quark masses}
\setcounter{equation}{0}

The quark masses are some of basic parameters of the standard
model. There are various hadronic phenomenology relating to the
light quark($u,\;d,\;s$) masses. For instance, they break the
chiral symmetry of QCD explicitly. $m_u-m_d$ breaks isospin
symmetry or charge symmetry, and $(m_u+m_d)/2-m_s$ breaks SU(3)
symmetry in hadron physics respectively. However, in QCD the
masses of the light quarks are not directly measurable in
inertial experiments, but enter the theory only indirectly as
parameters in the fundamental lagrangian. Therefore, it was an
active subject to determine light quark masses via
phenomenological method.

At low energy, the information about the light quark mass ratios can be
extracted in order by order by a rigorous, semiphenomenological method,
chiral perturbative theory(ChPT)\cite{GL85a,Wein79,GL82}. To the first
order, the results, $m_s/m_d=19$ and $m_u/m_d=0.556$, are
well-known\cite{GMO68}. Many authors have studied the mass ratios up to
next to leading order of the chiral expansion. Gasser and Leutwyler first
obtained $m_s/m_d=20.2$ and $m_u/m_d=0.554$\cite{GL85a}. Then Kaplan and
Manohar extracted $m_s/m_d=$15 to 23 and $m_u/m_d=$0 to 0.8 with very
larger error bar\cite{Kaplan86}. These values have been improved to
$m_s/m_d=(20.5\pm 2.5),\;m_u/m_d=0.52\pm 0.13$ by
Leutwyler\cite{Leutwyler90}, to $m_s/m_d=18,\;m_u/m_d=0.66$ by
Gerard\cite{Gerard90}, and to $m_s/m_d=21,\;m_u/m_d=0.30\pm 0.07$ by
Donoghue {\sl et al.}\cite{Donoghue92} respectively. Finally, Leutwyler
analysis previous results and obtained $m_s/m_d=18.9\pm
0.8,\;m_u/m_d=0.553\pm 0.043$\cite{Leut96}.

Although ChPT provides many information on light quark masses, it
is still not enough for further studies on vector meson physics:
At chiral limit, the unitarity implies that EMG model can not
predict the physics when energy $\sqrt{q^2}>2m\simeq 980$MeV. In
the other words, to study $\phi(1020)$ physics at chiral limit is
not consistent. However, if we take nonzero light quark masses,
EMG model will be unitarity when energy $\sqrt{q^2}>2(m+m_s)$. It
means that all light flavour vector meson resonances can be
included in to EMG model consistently. Hence, the strange quark
mass plays important role at $K^*$ and $\phi$ physics. Meanwhile,
ChPT can only provide information on light quark mass ratio. For
obtaining the individual quark masses, other approaches, such as
QCD sum rules\cite{Sum1,Sum2} or lattice
calculation\cite{Lattice} are needed. Some authors have pointed
out that the determination on the individual quark masses is
model-dependent\cite{GL82}. Thus for studying $K^*$ and $\phi$
physics in formalism of EMG model, it is better to determine light
quark masses by this model itself.

The Kaplan-Manohar ambiguity of ChPT has caused many debates. In
particular, due to this ambiguity, the authors of ref.\cite{Choi88} argued
that the observed mass spectrum is consistent with a broad range of quark
mass ratios, which specially includes the possibility $m_u=0$. However,
it is disagreed by other anthors\cite{Leut90,Donoghue92}. This problem
will also be discussed in EMG model. From lagrangian~(\ref{2.22}), we can
see that light current quark masses are defined uniquely in this model,
that they are just renormalized ``physical'' masses of u, d, and s quarks.
In principle, therefore, there is no Kaplan-Manohar ambiguity in EMG
model. Light current quark masses can be determined uniquely via meson
spectrum.

In general, in EMG model the light quark masses can be extracted
not only by pseudoscalar meson spectrum, but also by lowest
vector meson resonance spectrum. However, one-loop effects of
mesons will contribute to vector meson masses. As shown in the
above sections, the calculations to every vector meson resonance
are very complicate. It makes that the relationship between
vector meson spectrum and the light quark masses are also very
complicate and indirect. Thus here we will extract information
about the light current quark masses from pseudoscalar meson
spectrum and their decay constants. The vector meson spectrum
will be predicted by this formalism in other paper.

For the purpose of this paper, the lagrangian(~\ref{2.22}) can be
rewritten as follow
\begin{eqnarray}\label{9.1}
{\cal L}_{\chi}(x)={\cal L}_q(x)+{\cal L}_2^{(0)}(x)
  +{\cal L}[\pi(x),\eta_8(x)]+{\cal L}[K^\pm(x)]+{\cal L}[K^0(x)],
\end{eqnarray}
where ${\cal L}_q$ and ${\cal L}_2^{(0)}$ are free field lagrangian of
constituent quarks of pseudoscalar meson respectively,
\begin{eqnarray}\label{9.2}
{\cal L}_q&=&\sum_{i=u,d,s}\bar{q}_i(i\sla{\pa}-\bar{m}_i)q_i,
  \hspace{1in} \bar{m}_i=m+m_i, \nonumber \\
{\cal L}_2^{(0)}&=&\frac{F^2}{16}<\nabla_\mu U\nabla^\mu U^{\dag}>
               \nonumber \\
&=&\frac{F^2}{8}{\pa_\mu\Phi^a\pa^\mu\Phi^a+4a_\mu^a\pa^\mu\Phi^a+\cdots},
  \hspace{0.5in}a=1,2,\cdots,8.
\end{eqnarray}
${\cal L}[\Phi(x)]$ denotes quark-meson coupling lagrangian,
\begin{eqnarray}\label{9.3}
{\cal L}[\pi,\eta_8]&=&\frac{g_A}{\sqrt{2}}[(\pa_\mu\pi^++2a_\mu^+)
   \bar{u}\gamma^\mu\gamma_5d+c.c.]+\frac{g_A}{2}\sum_{i=u,d,s}
   (\pa_\mu P_i+2A_{\mu}^{(i)})\bar{q}_i\gamma^\mu\gamma_5q_i
     \nonumber \\
   &&+\frac{i}{\sqrt{2}}\kappa(m_u+m_d)(\pi^+\bar{u}\gamma_5d+c.c.)
  +i\kappa\sum_{i=u,d,s}m_iP_i\bar{q}_i\gamma_5q_i \nonumber \\
   &&+\frac{1}{2}(m_u+m_d)\pi^+\pi^-(\bar{u}u+\bar{d}d)
     +\frac{1}{2}\sum_{i=u,d,s}m_iP_i^2\bar{q}_iq_i,\\
{\cal L}[K^\pm]&=&\frac{g_A}{\sqrt{2}}[(\pa_\mu K^++2A_\mu^+)
   \bar{u}\gamma^\mu\gamma_5s+c.c.]
   +\frac{i}{\sqrt{2}}\kappa(m_u+m_s)(K^+\bar{u}\gamma_5s+c.c.)
      \nonumber \\
  &&+\frac{1}{2}(m_u+m_s)K^+K^-(\bar{u}u+\bar{s}s),\nonumber \\
{\cal L}[K^0]&=&\frac{g_A}{\sqrt{2}}[(\pa_\mu K^0+2A_\mu^0)
   \bar{d}\gamma^\mu\gamma_5s+(\pa_\mu\bar{K}^0+2\bar{A}_\mu^0)
   d\gamma^\mu\gamma_5\bar{s}]        \nonumber \\
   &&+\frac{i}{\sqrt{2}}\kappa(m_d+m_s)(K^0\bar{d}\gamma_5s
  +\bar{K}^0d\gamma_5\bar{s})
  +\frac{1}{2}(m_d+m_s)K^0\bar{K}^0(\bar{d}d+\bar{s}s),\nonumber
\end{eqnarray}
where $c.c.$ denotes charge conjugate term of previous term,
\begin{eqnarray}\label{9.4}
P_u=\pi_3+\frac{1}{\sqrt{3}}\eta_8,\;\;\;\;\;\;\;\;
P_d=-\pi_3+\frac{1}{\sqrt{3}}\eta_8,\;\;\;\;\;\;\;\;
P_s=-\frac{2}{\sqrt{3}}\eta_8,
\end{eqnarray}
and $a_\mu^a$, $A_{\mu}^{(u)},\;A_{\mu}^{(d)}$ and
$A_{\mu}^{(s)}$ etc. are axial-vector external fields
corresponding these pseudoscalar meson fields. From
eq.~(\ref{9.3}) we can see that the $K^\pm$-quark coupling and
$K^0$-quark coupling are similar to $\pi^\pm$-quark coupling.
Thus we only need to calculate masses and decay constants of
$\pi^\pm,\;\pi^0$ and $\eta^8$. Then masses and decay constants
of $K^\pm$ can be obtained via replacing $m_d$ by $m_s$ in one of
$\pi^\pm$, and masses and decay constants of $K^0$ can be
obtained via replacing $m_u$ by $m_s$ in one of $\pi^\pm$.

At large $N_c$ limit, the kinetic terms of pseudoscalar mesons
can be obtained via calculating one-point and two-point effective
action. Note that here the light quark masses will not be
expanded perturbatively, but enter quark propagator and be
evaluated in non-perturbative method. Explicitly, the kinetic
term of charge pion fields reads
\begin{eqnarray}\label{9.5}
{\cal L}_2[\pi^\pm(x)]&=&\frac{F^2(m_u,m_d)}{4}\pa_\mu\pi^+\pa^\mu\pi^-
  +\frac{\bar{f}^2(m_u,m_d)}{2}(a_\mu^+\pa^\mu\pi^-+c.c.)
  -\frac{F_0^2}{4}\bar{M}^2(m_u,m_d)\pi^+\pi^-\nonumber \\
 &&+\int\frac{d^4q}{(2\pi)^4}
 e^{iq\cdot x}\{\frac{\alpha(q^2;m_u,m_d)}{2}iq^\mu(a_\mu^+(x)\pi^-(q)
  +c.c.)-\frac{F_0^2}{4}\beta(q^2;m_u,m_d))\pi^+(q)\pi^-(x)\},
\end{eqnarray}
where
\begin{eqnarray}\label{9.6}
&&F^2(m_u,m_d)=F_0^2+\frac{3}{2}g^2g_A^2m_{ud}(2m+\bar{m}_{ud})
  +3g^2\kappa g_Am_{ud}\bar{m}_{ud}\nonumber \\
&&\;\;\;\;\;\;\;\;-\frac{N_c}{2\pi^2}g_A
 (g_A\bar{m}_{ud}+2\kappa m_{ud})\int_0^1dt[\bar{m}_u+t(m_d-m_u)]
\ln{(x_u^2+t(x_d^2-x_u^2))}\nonumber\\
&&\;\;\;\;\;\;\;\;
+4\kappa^2 m_{ud}^2\{(\frac{g^2}{4}-\frac{F_0^2B_0}{48m^3}
  -\frac{N_c}{24\pi^2})-\frac{N_c}{8\pi^2}\int_0^1dt\cdot t(1-t)
[3\ln{(x_u^2+t(x_d^2-x_u^2))}
-\frac{\bar{m}_u\bar{m}_d}{\bar{m}_u^2+t(\bar{m}_d^2-\bar{m}_u^2)}]\},
   \nonumber\\
&&\bar{f}^2(m_u,m_d)=F_0^2+\frac{3}{2}g^2g_A^2m_{ud}(4m+\bar{m}_{ud})
  +\frac{3}{2}g^2\kappa g_Am_{ud}\bar{m}_{ud}\nonumber \\
&&\;\;\;\;\;\;\;\;-\frac{N_c}{2\pi^2}g_A
 (g_A\bar{m}_{ud}+\kappa m_{ud}+m_d)\int_0^1dt[\bar{m}_u+t(m_d-m_u)]
\ln{(x_u^2+t(x_d^2-x_u^2))},\nonumber\\
&&\bar{M}^2(m_u,m_d)=B_0m_{ud}\{\frac{1}{2}(x_u^3+x_d^3)
 +\frac{N_c}{2\pi^2}\frac{m^3}{F_0^2B_0}(x_u^3\ln{x_u^2}+x_d^3\ln{x_d^3})
 -\frac{\kappa^2}{4}\frac{m_{ud}}{m}(x_u^2+x_d^2)\}\nonumber \\
&&\;\;\;\;\;\;\;\;+\frac{3}{2}g^2\kappa^2\frac{(m_d^2-m_u^2)^2}{F_0^2}
 -\frac{N_c}{2\pi^2}\kappa^2\frac{m_{ud}^2}{F_0^2}\int_0^1dt
 [2\bar{m}_u^2-\bar{m}_u\bar{m}_d+2t(\bar{m}_d^2-\bar{m}_u^2)]
 \ln{(x_u^2+t(x_d^2-x_u^2))},\\
&&\alpha(q^2;m_u,m_d)=-\frac{N_c}{2\pi^2}g_A
 (g_A\bar{m}_{ud}+\kappa m_{ud})\int_0^1dt[\bar{m}_u+t(m_d-m_u)]
\ln{\left(1-\frac{t(1-t)q^2}{\bar{m}_u^2+t(\bar{m}_d^2-\bar{m}_u^2)}
 \right)},\nonumber \\
&&\beta(q^2;m_u,m_d)=\frac{N_c}{2\pi^2}g_A\frac{q^2}{F_0^2}
 (g_A\bar{m}_{ud}+2\kappa m_{ud})\int_0^1dt[\bar{m}_u+t(m_d-m_u)]
\ln{\left(1-\frac{t(1-t)q^2}{\bar{m}_u^2+t(\bar{m}_d^2-\bar{m}_u^2)}
 \right)}\nonumber \\&&\;\;\;\;\;\;\;\;
-\frac{N_c}{2\pi^2}\kappa^2\frac{m_{ud}^2}{F_0^2}\int_0^1dt
\{[2\bar{m}_u^2-\bar{m}_u\bar{m}_d+2t(\bar{m}_d^2-\bar{m}_u^2)
  -3t(1-t)q^2]\ln{\left(1-\frac{t(1-t)q^2}{\bar{m}_u^2
  +t(\bar{m}_d^2-\bar{m}_u^2)}\right)}\nonumber \\&&\hspace{2in}
  +t(1-t)q^2\left(2-\frac{\bar{m}_u\bar{m}_d}{\bar{m}_u^2
  +t(\bar{m}_d^2-\bar{m}_u^2)}\right)\}, \nonumber
\end{eqnarray}
where $m_{ij}=m_i+m_j$ and $\bar{m}_{ij}=\bar{m}_i+\bar{m}_j
(i,j=u,d,s)$. It should be pointed out that $\alpha(q^2;m_u,m_d)$
is order $q^2$ at least and $\beta(q^2;m_u,m_d)$ is order $q^4$
at least. Since in this paper we focus on pseudoscalar meson
spectrum, these high order derivative terms should obey motion
equation of pseudoscalar mesons. In momentum space, the motion
equation of physical pseudoscalar mesons is generally written
\begin{eqnarray}\label{9.7}
(q^2-m_{\vphi}^2)\vphi(q)=-if_{\vphi}q^\mu A_\mu^{(\vphi)}(-q),
\end{eqnarray}
where $m_{\vphi}$ and $f_{\vphi}$ are physical mass and decay constants of
pseudoscalar, e.g., $m_\pi=135$MeV and $f_\pi=185.2$MeV. Due to this
motion equation, we have
\begin{eqnarray}\label{9.8}
\alpha(q^2;m_u,m_d)a_\mu^+(x)\pi^-(q)
&=&\alpha(m_{\pi}^2;m_u,m_d)a_\mu^+(x)\pi^-(q),\nonumber \\
\beta(q^2;m_u,m_d)\pi^+(q)\pi^-(-q)
&=&\beta(m_\pi^2;m_u,m_d)\pi^+(q)\pi^-(-q)-\frac{i}{2}q^\mu f_\pi
  \beta'(m_{\pi}^2;m_u,m_d)(a_\mu^+(x)\pi^-(q)+c.c.),
\end{eqnarray}
where
\begin{eqnarray}\label{9.9}
\beta'(m_{\pi}^2;m_u,m_d)=\frac{d}{dq^2}
 \beta'(q^2;m_u,m_d)|_{q^2=m_{\pi}^2}.
\end{eqnarray}
Thus eq.(~\ref{3.15}) can be rewritten
\begin{eqnarray}\label{9.10}
{\cal L}_2[\pi^\pm(x)]&=&\frac{F^2(m_u,m_d)}{4}\pa_\mu\pi^+\pa^\mu\pi^-
-\frac{F_0^2}{4}\{\bar{M}^2(m_u,m_d)+\beta(m_{\pi}^2;m_u,m_d)\}\pi^+\pi^-
   \nonumber \\
&&+\{\frac{\bar{f}^2(m_u,m_d)}{2}+\frac{\alpha(m_\pi^2;m_u,m_d)}{2}
   +\frac{F_0^2}{4F(m_u,m_d)}f_\pi\beta'(m_{\pi}^2;m_u,m_d)\}
  (a_\mu^+\pa^\mu\pi^-+c.c.).
\end{eqnarray}

The kinetic term of $\pi^0$ and $\eta_8$ reads
\begin{eqnarray}\label{9.11}
{\cal L}_2[\pi^0(x),\eta_8(x)]=\sum_{i=u,d,s}\{\frac{F_i^2(m_i)}{8}
  \pa_\mu P_i\pa^\mu P_i-\frac{F_0^2}{8}\bar{M}_i^2(m_i)P_i^2
  -\frac{F_0^2}{8}\int\frac{d^4q}{(2\pi)^4}e^{iq\cdot x}
  \beta_i(q^2;m_i)P_i(q)P_i(x)\},
\end{eqnarray}
where
\begin{eqnarray}\label{9.12}
F_i^2(m_i)&=&\frac{F_0^2}{2}+3g^2g_A^2m_i(2m+m_i)+6g^2\kappa g_Am_i
  \bar{m}_i+8\kappa^2(\frac{g^2}{4}-\frac{F_0^2B_0}{48m^3}
  -\frac{N_c}{48\pi^2})m_i^2\nonumber \\
&&-\frac{N_c}{2\pi^2}(g_A^2\bar{m}_i^2+2g_A\kappa m_i\bar{m}_i
   +\kappa^2 m_i^2)\ln{x_i^2}, \nonumber \\
\bar{M}_i^2(m_i)&=&(B_0+\frac{N_c}{\pi^2}\frac{m^3}{F_0^2}\ln{x_i^2})
    m_ix_i^2(x_i-\frac{m_i}{m}\kappa^2), \\
\beta_i(q^2;m_i)&=&\frac{N_c}{2\pi^2F_0^2}q^2\{\frac{\kappa^2}{3}m_i^2
 +\left(g_A^2\bar{m}_i^2+2\kappa m_i\bar{m}_i+2\kappa^2 m_i^2
  [\bar{m}_i^2q^{-2}-3t(1-t)]\right)\int_0^1dt
  \ln{(1-\frac{t(1-t)q^2}{\bar{m}_i^2})}\}.  \nonumber
\end{eqnarray}
Since the decay constants for neutral mesons cannot be extracted
directly from the data. It means that the decay constants for
neutral mesons can not be used to determine light current quark
masses. Therefore, in this paper we do not need to evaluate the
decay constants for neutral pion and $\eta_8$. In addition,
because auxiliary fields $P_i$ do not lie in physical hadron
spectrum, the equation of motion (\ref{9.7}) can not be used in
(\ref{9.11}) simply. We will use propagator method to deal with
the terms with high power momenta in (\ref{9.11}) and diagonalize
$\pi_3-\eta_8$ mixing.

Next we evaluate one-loop effects of pseudoscalar mesons. Due to
parity conservation, there are only tadpole diagrams of
pseudoscalar mesons contributing to masses of decay constants of
$0^-$ mesons(fig. 1). Moreover, we can expect that, in mass
spectrum and decay constants of pseudoscalar mesons, the dominant
one-loop effects is generated by the lowest order effective
lagrangian, because neglect is $O(m_q^3)$ and suppressed by
$N_c^{-1}$ expansion. It is fortunate that the formalism is
renormalized up to this order.

The lowest order effective lagrangian is well-known
\begin{eqnarray}\label{9.13}
{\cal L}_2&=&\frac{F_0^2}{16}<\nabla_\mu U\nabla^\mu U^{\dag}
  +\chi U^{\dag}+U\chi^{\dag}> \nonumber \\
  &=&\frac{F_0^2}{48}<[\lambda^a,\Delta_\mu][\lambda^a,\Delta^\mu]>
    +\frac{3F_0^2}{256}<\lambda^a\lambda^a(\xi\chi^{\dag}\xi
     +\xi^{\dag}\chi\xi^{\dag})>,
\end{eqnarray}
Substituting expansion(~\ref{4.1}) into ${\cal L}_2$ and retaining terms
up to and including $\vphi^2$ one obtains
\begin{eqnarray}\label{9.14}
{\cal L}_2\rightarrow \bar{\cal L}_2+\frac{F_0^2}{8}(\pa_\mu\vphi^a
 \pa^\mu\vphi^a-m_\vphi^2\vphi^a\vphi^a)-\frac{F_0^2}{16}
 <[\vphi,\Delta_\mu][\vphi,\Delta^\mu]+\frac{1}{4}\{\vphi,\vphi\}
 (\bar{\xi}\chi^{\dag}\bar{\xi}+\bar{\xi}^{\dag}\chi\bar{\xi}^{\dag}
 -2{\cal M})>,
\end{eqnarray}
where we have omitted some terms which do not contribute to masses and
decay constants via one-loop graphs. The contribution of tadpole graphs
can be calculated easily
\begin{eqnarray}\label{9.15}
{\cal L}_2^{(\rm tad)}&=&-\frac{1}{4}
 <[\lambda^a,\Delta_\mu][\lambda^a,\Delta^\mu]+\frac{1}{4}
 \{\lambda^a,\lambda^a\}(\bar{\xi}\chi^{\dag}\bar{\xi}
  +\bar{\xi}^{\dag}\chi\bar{\xi}^{\dag}-2{\cal M})>
 \int\frac{d^4k}{(2\pi)^4}\frac{i}{k^2-m_\vphi^2+i\ep} \nonumber \\
&=&\frac{1}{4}(m_\vphi^2N_{\ep}-\frac{m_\vphi^2}{16\pi^2}
  \ln{\frac{m_\vphi^2}{\mu^2}})
  <[\lambda^a,\Delta_\mu][\lambda^a,\Delta^\mu]+\frac{1}{4}
 \{\lambda^a,\lambda^a\}(\bar{\xi}\chi^{\dag}\bar{\xi}
  +\bar{\xi}^{\dag}\chi\bar{\xi}^{\dag}-2{\cal M})>,
\end{eqnarray}
where
$$N_{\ep}=\frac{1}{16\pi^2}\{\frac{2}{\ep}+\ln{(4\pi)}+\gamma+1\}.$$
Comparing eq.(~\ref{9.15}) and eq.(~\ref{9.12}) we can see that the
divergence $N_{\ep}$ can be absorbed by free parameters $F_0$ and $B_0$.
Thus the sum of tree graphs and tadpole contribution is
\begin{eqnarray}\label{9.16}
{\cal L}_2^{(t)}=\frac{F_0^2}{48}(1-3\mu_\vphi)
v <[\lambda^a,\Delta_\mu][\lambda^a,\Delta^\mu]>
 +\frac{3F_0^2}{256}(1-\frac{8}{3}\mu_\vphi)<\lambda^a\lambda^a
  (\xi\chi^{\dag}\xi+\xi^{\dag}\chi\xi^{\dag})>,
\end{eqnarray}
where
\begin{eqnarray}\label{9.17}
\mu_\vphi=\frac{m_\vphi^2}{4\pi^2F_0^2}\ln{\frac{m_\vphi^2}{\mu^2}}.
\end{eqnarray}
Here we appoint that $m_\vphi=m_\pi$ when $a=1,2,3$,
$m_\vphi=m_{_K}$ when $a=4,5,6,7$ and $m_\vphi=m_{\eta_8}$ when $a=8$.

For extracting $(m_{_{K^+}})_{\rm QCD}$ from experimental data,
the electromagnetic mass splitting of $K$-meson is required. The
prediction of Dashen theorem\cite{Dashen69},
$(m_{_{K^+}}-m_{_{K^0}})_{e.m.}=1.3$MeV, has been corrected in
several recent analysis with considering contribution from vector
meson exchange. A larger correction is first obtained by
Donoghue, Holstein and Wyler\cite{DHW93}, who find
$(m_{_{K^+}}-m_{_{K^0}})_{e.m.}=2.3$MeV. Then Bijnens and
Prades\cite{Bijnens97}, who evaluated both long-distance
contribution using ENJL model and short-distance contribution
using perturbative QCD and factorization, find
$(m_{_{K^+}}-m_{_{K^0}})_{e.m.}=2.4\pm 0.3$MeV at $\mu=m_\rho$.
Gao {\sl et.al.}\cite{Gao97} also gave
$(m_{_{K^+}}-m_{_{K^0}})_{e.m.}=2.5$MeV. Baur and
Urech\cite{BU96}, however, obtained a smaller correction,
$(m_{_{K^+}}-m_{_{K^0}})_{e.m.}=1.6$MeV at $\mu=m_\rho$. In
addition, calculation of lattice QCD\cite{Duncan96} found
$(m_{_{K^+}}-m_{_{K^0}})_{e.m.}=1.9$MeV. These estimates indicate
that the corrections to Dashen theorem are indeed substantial. In
this the present paper, we average the above results and take
$(m_{_{K^+}}-m_{_{K^0}})_{e.m.}=2.1\pm 0.1$MeV at energy scale
$\mu=m_\rho$.

From eqs.~(\ref{9.10}) and (\ref{9.16}), the masses and decay
constants of koan and charge pion can be obtained via solve the
following equations
\begin{eqnarray}\label{9.18}
m_{\pi^+}^2&=&\frac{F_0^2}{F_{_R}^2(m_u,m_d)}\{\bar{M}_{_R}^2(m_u,m_d)
  +\beta(m_\pi^2;m_u,m_d)\}
   \nonumber \\
m_{_{K^+}}^2&=&\frac{F_0^2}{F_{_R}^2(m_u,m_s)}\{\bar{M}_{_R}^2(m_u,m_s)
  +\beta(m_{_K}^2;m_u,m_s)\},
   \nonumber \\
m_{_{K^0}}^2&=&\frac{F_0^2}{F_{_R}^2(m_d,m_s)}\{\bar{M}_{_R}^2(m_d,m_s)
  +\beta(m_{_K}^2;m_d,m_s)\}, \\
f_{\pi^+}&=&\frac{\bar{f}_{_R}^2(m_u,m_d)}{F_{_R}(m_u,m_d)}+
  \frac{\alpha(m_\pi^2;m_u,m_d)}{F_{_R}(m_u,m_d)}
  +\frac{F_0^2}{2F_{_R}^2(m_u,m_d)}f_{\pi^+}\beta'(m_\pi^2;m_u,m_d),
   \nonumber \\
f_{_{K^+}}&=&\frac{\bar{f}_{_R}^2(m_u,m_s)}{F_{_R}(m_u,m_s)}+
  \frac{\alpha(m_{_K}^2;m_u,m_s)}{F_{_R}(m_u,m_s)}
  +\frac{F_0^2}{2F_{_R}^2(m_u,m_s)}f_{_K}\beta'(m_{_K}^2;m_u,m_s),
    \nonumber \\
f_{_{K^0}}&=&\frac{\bar{f}_{_R}^2(m_d,m_s)}{F_{_R}(m_d,m_s)}+
  \frac{\alpha(m_{_K}^2;m_d,m_s)}{F_{_R}(m_d,m_s)}
  +\frac{F_0^2}{2F_{_R}^2(m_d,m_s)}f_{_K}\beta'(m_{_K}^2;m_d,m_s),
  \nonumber
\end{eqnarray}
where subscript ``R'' denotes renormalized quantity,
\begin{eqnarray}\label{9.19}
F_{_R}^2(m_u,m_d)&=&F^2(m_u,m_d)-F_0^2(2\mu_\pi+\mu_{_K}), \nonumber \\
F_{_R}^2(m_i,m_s)&=&F^2(m_i,m_s)-\frac{3}{4}F_0^2(\mu_\pi+2\mu_{_K}
    +\mu_{\eta_8}), \hspace{1in} (i=u,d)\nonumber \\
\bar{M}_{_R}^2(m_u,m_d)&=&\bar{M}^2(m_u,m_d)
  -B_0(m_u+m_d)(\frac{3}{2}\mu_\pi
  +\mu_{_K}+\frac{1}{6}\mu_{\eta_8}), \nonumber \\
\bar{M}_{_R}^2(m_i,m_s)&=&\bar{M}^2(m_i,m_s)-B_0(m_i+m_s)
  (\frac{3}{4}\mu_\pi+\frac{3}{2}\mu_{_K}+\frac{5}{12}\mu_{\eta_8}), \\
\bar{f}_{_R}^2(m_u,m_d)&=&\bar{f}^2(m_u,m_d)-F_0^2(2\mu_\pi+\mu_{_K}),
  \nonumber \\
\bar{f}_{_R}^2(m_i,m_s)&=&\bar{f}^2(m_i,m_s)-
   \frac{3}{4}F_0^2(\mu_\pi+2\mu_{_K}+\mu_{\eta_8}), \nonumber
\end{eqnarray}
Here the quantity $\mu_\vphi$ depends on the renormalization scale $\mu$.
It has been recognized that the scale plays an important role in
low-energy QCD. In this formalism, many parameters, such as light current
quark masses, constituent quark mass $m$, axial-vector coupling constant
$g_A$, are scale-dependent. The characteristic scale of the model is
described by the universal coupling constant $g$ which is determined by
the first KSRF sum rule at energy scale $\mu=m_\rho$. Hence we take
$\mu=m_\rho=770$MeV in $\mu_\vphi$, and the physical masses and decay
constants, however, should be independent of
the renormalization scale.

In order to obtain the masses of $\pi^0$ and $\eta^8$, the $\pi_3-\eta_8$
mixing in eq.~(\ref{9.11}) must be diagonalized. Eq.(\ref{9.11})
together with eq.(\ref{9.16}) lead to the quadratic terms for the
$\pi_3$ and $\eta_8$ are of the form
\begin{eqnarray}\label{9.20}
S_2[\pi_3,\eta_8]=\int\frac{d^4q}{(2\pi)^4}\{
  \frac{1}{2}(q^2-M_3^2(q^2))\pi_3^2
  +\frac{1}{2}(q^2-M_8^2(q^2))\eta_8^2-M_{38}^2(q^2)\pi_3\eta_8\}
\end{eqnarray}
where
\begin{eqnarray}\label{9.21}
M_3^2(q^2)&=&\frac{F_0^2}{F_3^2}\{\bar{M}_u^2(m_u)+\bar{M}_d^2(m_d)
   +\beta_u(q^2;m_u)+\beta_d(q^2;m_d)-2B_0\hat{m}
   (\frac{3}{2}\mu_\pi+\mu_{_K}+\frac{1}{6}\mu_{\eta_8})\}, \nonumber \\
M_8^2(q^2)&=&\frac{F_0^2}{3F_8^2}\{\bar{M}_u^2(m_u)+
  \bar{M}_d^2(m_d)+4\bar{M}_s^2(m_s)+\beta_u(q^2;m_u)+\beta_d(q^2;m_d)
  +4\beta_s(q^2;m_s) \nonumber \\ &&-2B_0(\hat{m}+2m_s)
  (2\mu_{_K}+\frac{2}{3}\mu_{\eta_8})-2B_0\hat{m}(-\frac{3}{2}\mu_\pi
  +\mu_{_K}+\frac{1}{2}\mu_{\eta_8})\},  \\
M_{38}^2(q^2)&=&\frac{F_0^2}{\sqrt{3}F_3F_8}\{\bar{M}_u^2(m_u)
   -\bar{M}_d^2(m_d)+\beta_u(q^2;m_u)-\beta_d(q^2;m_d)-B_0(m_u-m_d)
   (\frac{3}{2}\mu_\pi+\mu_{_K}+\frac{1}{6}\mu_{\eta_8})\}\nonumber \\
  &&+q^2\frac{F_u^2(m_u)-F_d^2(m_d)}{\sqrt{3}F_3F_8},\nonumber
\end{eqnarray}
with $\hat{m}=(m_u+m_d)/2$ and
\begin{eqnarray}\label{9.22}
F_3^2&=&F_u^2(m_u)+F_d^2(m_d)-F_0^2(2\mu_\pi+\mu_{_K}),\nonumber \\
F_8^2&=&\frac{1}{3}\{F_u^2(m_u)+F_d^2(m_d)+4F_s^2(m_s)\}
    -3F_0^2\mu_{_K}.
\end{eqnarray}

Due to $\pi_3-\eta_8$ mixing, the ``physical'' propagators of $\pi^0$ and
$\eta_8$ are obtained via the chain approximation in momentum space
\begin{eqnarray}\label{9.23}
\frac{i}{q^2-m_{\pi^0}^2+i\ep}&=&\frac{i}{q^2-M_3^2(q^2)+i\ep}
  +\frac{iM_{38}^4(q^2)}{(q^2-M_8^2(q^2)+i\ep)(q^2-M_3^2(q^2)+i\ep)^2}
  +\cdots \nonumber \\
 &=&\frac{i}{q^2-M_3^2(q^2)-\frac{M_{38}^4(q^2)}{q^2-M_8^2(q^2)}+i\ep},
     \nonumber \\
\frac{i}{q^2-m_{\eta_8}^2+i\ep}&=&\frac{i}{q^2-M_8^2(q^2)+i\ep}
  +\frac{iM_{38}^4(q^2)}{(q^2-M_3^2(q^2)+i\ep)(q^2-M_8^2(q^2)+i\ep)^2}
  +\cdots \nonumber \\
 &=&\frac{i}{q^2-M_8^2(q^2)-\frac{M_{38}^4(q^2)}{q^2-M_3^2(q^2)}+i\ep}.
\end{eqnarray}
Then the masses of $\pi^0$ and $\eta_8$ are just solutions of the
following equations,
\begin{eqnarray}\label{9.24}
m_{\pi^0}^2&=&M_3^2(m_{\pi^0}^2)+\frac{M_{38}^2(m_{\pi^0}^2)}
   {m_{\pi^0}^2-M_8^2(m_{\pi^0}^2)}, \nonumber \\
m_{\eta^8}^2&=&M_8^2(m_{\eta_8}^2)+\frac{M_{38}^2(m_{\eta_8}^2)}
   {m_{\eta_8}^2-M_8^2(m_{\eta_8}^2)}.
\end{eqnarray}

In the above results, the parameters $\kappa$, $F_0$ and
$B_0$ are still not determined. In order to determine them and three light
quark masses, six inputs are required. In this paper we choose
$f_{\pi^+}=185.2\pm 0.5$MeV, $f_{_{K^+}}=226.0\pm 2.5$MeV,
$m_{\pi^0}=134.98$MeV, $m_{_{K^0}}=497.67$MeV and $(m_{_{K^+}})_{\rm
QCD}=491.6\pm 0.1$MeV. Another input is $m_d-m_u=3.9\pm 0.22$MeV, which
is extracted from $\omega\rightarrow\pi^+\pi^-$ decay at energy scale
$\mu=m_\rho$. Recalling $m=480$MeV, $g_A=0.75$ and $g=\pi^{-1}$ for
$N_c=3$, we can fit light quark masses as in table 1.

In table 1, the errors of results are from uncertainties in decay constant
of $K^+$. electromagnetic mass splitting of $K$-mesons and isospin
violation parameter $m_d-m_u$ respectively. The first column in table 1 we
show the results for $f_{_{K^+}}=223.5$MeV.  The third column corresponds
the center value of $f_{_{K^+}}$, $226.0$MeV and the fifth column
corresponds $f_{_{K^+}}=228.5$MeV. Then from table 1 we have
\begin{eqnarray}\label{9.25}
m_s&=&160\pm 15.5{\rm MeV},\hspace{0.5in}
m_d=7.9\pm 2.7{\rm MeV},\hspace{0.5in}
m_u=4.1\pm 1.5{\rm MeV},\nonumber \\
\frac{m_s}{m_d}&=&20.2\pm 3.0,\hspace{0.8in}
\frac{m_u}{m_d}=0.5\pm 0.09.
\end{eqnarray}
Here the large errors are from the uncertainty of $f_{_{K^+}}$.
We can see that our results agree with one obtained by other
various approaches(e.g., QCD sum rule or ChPT) well. From table 1
we also have $(m_{\pi^+}-m_{\pi^0})_{\rm QCD}=-0.25$MeV (where
the contribution from $\pi_3-\eta'$ mixing is neglected). This
result allows that the electromagnetic mass splitting of pion is
$(m_{\pi^+}-m_{\pi^0})_{\rm e.m.}=4.8$MeV.

Finally, it is interesting to expand non-perturtation results (\ref{5.1})
up to the next to leading order of the chiral expansion and compare with
one of ChPT. If we neglect the mass difference $m_d-m_u$, up to this order
the masses of decay constants of koan and pion read
\begin{eqnarray}\label{9.26}
m_\pi^2&=&B_0(m_u+m_d)\{1+\frac{1}{2}\mu_\pi-\frac{1}{6}\mu_{\eta_8}
   +[\frac{1}{2}(3-\kappa^2)+\frac{3m^2}{\pi^2F_0^2}
     (\frac{m}{B_0}-\kappa g_A-\frac{g_A^2}{2}-\frac{B_0}{6m}g_A^2)]
   \frac{m_u+m_d}{m}\}, \nonumber \\
m_{_K}^2&=&B_0(\hat{m}+m_s)\{1+\frac{1}{3}\mu_{\eta_8}
   +[\frac{1}{2}(3-\kappa^2)+\frac{3m^2}{\pi^2F_0^2}
     (\frac{m}{B_0}-\kappa g_A-\frac{g_A^2}{2}-\frac{B_0}{6m}g_A^2)]
   \frac{\hat{m}+m_s}{m}\}, \\
f_\pi&=&F_0\{1-\mu_\pi-\frac{1}{2}\mu_{_K}+\frac{3}{2\pi^2}g_A^2
   \frac{m(m_u+m_d)}{F_0^2}\}, \nonumber \\
f_{_K}&=&F_0\{1-\frac{3}{8}\mu_\pi-\frac{3}{4}\mu_{_K}
   -\frac{3}{8}\mu_{\eta_8}+\frac{3}{2\pi^2}g_A^2
   \frac{m(\hat{m}+m_s)}{F_0^2}\}. \nonumber
\end{eqnarray}
Comparing the above equation with one of ChPT\cite{GL85a}, we predict the
$O(p^4)$ chiral coupling constants, $L_4,\;L_5\;L_6$ and $L_8$, as follow
\begin{eqnarray}\label{9.27}
L_4&=&L_6=0, \hspace{1in}L_5=\frac{3m}{32\pi^2B_0}g_A^2, \nonumber \\
L_8&=&\frac{F_0^2}{128B_0m}(3-\kappa^2)+\frac{3m}{64\pi^2B_0}
  (\frac{m}{B_0}-\kappa g_A-\frac{g_A^2}{2}-\frac{B_0}{6m}g_A^2)
  +\frac{L_5}{2}.
\end{eqnarray}
Numerically, inputting $\kappa=0.35\pm 0.15$, $F_0=0.156$GeV and
$B_0=1.7\pm 0.4$GeV, we obtain $L_5=(1.2\pm 0.3)\times 10^{-3}$
and $L_8=(0.0.47\pm 0.25)\times 10^{-3}$. These value agree with
one of ChPT, $L_5=(1.4\pm 0.5)\times 10^{-3}$ and $L_8=(0.47\pm
0.3)\times 10^{-3}$ at energy scale $\mu=m_\rho$. This result
indicates that the contribution from scalar meson resonance
exchange is small. In fact, in hadron spectrum, there is no
scalar meson octet or singlet which belong to composite fields of
$q\bar{q}$. Thus it is a {\it ad hoc} assumption to argue some
low energy coupling constants, such as $L_5$ and $L_8$, receiving
large contribution from scalar meson exchange.

\begin{table}[hpb]
\centering
 \begin{tabular}{cccccc}
        &Fit 1&Fit 2&Fit 3& Fit 4&Fit 5 \\ \hline
$^{\dag}f_{\pi^+}$&185.2&185.2&185.2&185.2&185.2 \\
$^{\dag}f_{_{K^+}}$&223.5&224.1&226.0&227.8&228.5 \\
$^{\dag}m_{\pi^0}$&134.98&134.98&134.98&134.98&134.98 \\
$^{\dag}m_{_{K^0}}$&497.67&497.67&497.67&497.67&497.67 \\
$^{\dag}(m_{_{K^+}})_{\rm QCD}$&$491.6\pm 0.1$&$491.6\pm 0.1$&
  $491.6\pm 0.1$&$491.6\pm 0.1$&$491.6\pm 0.1$ \\
$^{\dag}B(\omega\rightarrow\pi\pi)$&$1.95\%$&$1.95\%$&$(2.11\pm 0.20)\%$
  &$(2.21\pm 0.25)\%$&$(2.21\pm 0.30)\%$ \\
$m_s$&144.2&150.0&161.8 &172.0&175.4 \\
$m_d$&$6.26$&$6.83$&$7.94\pm 0.07$&$8.94\pm 0.1$&
      $9.53\pm 0.11$ \\
$m_u$&$2.56$&$3.13$&$4.10\mp 0.07$&$5.03\mp 0.1$&
      $5.63\mp 0.11$ \\
$m_u+m_d$&8.92&9.96&12.04&13.97&15.16 \\
$m_d-m_u$&$3.7$&$3.7$&$3.84\pm 0.14$&$3.9\pm 0.2$&$3.9\pm 0.22$ \\
$f_{_{K^0}}$&224.0&224.7&226.6&228.4&229.0 \\
$(m_{\pi^+})_{\rm QCD}$&134.74&134.74&134.73&134.73&134.73 \\
$m_{\eta_8}$&608.8&616.9&628.8&636.8&641.0 \\
$\kappa$&0.5&0.4&0.3&0.25&0.2 \\
$F_0$&156.95&156.7&156.1&155.6&155.3 \\
$B_0$&2141.1&1886.6&1551.1&1328.6&1216.8 \\
\end{tabular}
\begin{minipage}{5in}
\caption{Light current quark masses predicted by the masses and decay
constants of pseudoscalar mesons at energy scale $\mu=m_\rho$. Here
$\kappa$ is dimensionless, other dimensionful quantities are in MeV, and
$``\dag''$ denotes input.}
\end{minipage}
\end{table}

\section{Summary and Discussion}
\setcounter{equation}{0}

This paper is motivated by two purposes: Theoretically,
convergence of chiral expansion at vector meson energy scale
should be studied. Such a study is important, since it determines
whether a well-defined chiral theory can be constructed in this
energy region. Phenomenologically, we hope to study vector meson
physics in more rigorous way. In the other words, we hope that
these studies are not limited at the leading order of chiral
expansion. For achieving these two purposes, we need first choose
a model, which should satisfy the following requirements: 1) This
model must yield convergent chiral expansion at vector meson
energy scale. 2) Unitarity of $S$-matrix in this model must be
insured. 3) This model must be available to evaluate high order
contributions of low energy expansion. 4) The theoretical
predictions can be derived practically, and the results match
with data. Obviously, the ChPT-like
models\cite{Brise97,Eck89,Bando} do not satisfy the third
requirement, since too many free parameters are needed as
perturbative order raising. In addition, the Extend NJL-like
models\cite{Esp90,Chan85,ENJL} do not satisfy the first and the
second requirements at vector meson energy scale. Therefore, EMG
model, which satisfy all above requirements, is the simplest and
practical model for achieving the purposes of this paper.

There are usually three perturbative expansions in energy below
CSSB scale. They are external momentum expansion, light quark
mass expansion and $N_c^{-1}$ expansion. It is well-known that
the light quark mass expansion and $N_c^{-1}$ expansion are
well-defined for this whole energy region. In the other words, in
these two expansion, the dominant contribution is from the
leading order of perturbative expansion and the calculation for
this order is a controlled approximation in that error bars can
be put in the predictions. This point has been explicitly
confirmed in this paper. For example, the direct calculation show
that high order contribution of $N_c^{-1}$ expansion is indeed
about $\Gamma_\rho:m_\rho$ which is around $20\%$. However, the
external momentum expansion is not similar to other two
expansions. According to usual argument on external momentum
expansion, this expansion should be in powers of
$p^2/\Lambda_\chi^2$. For $p^2\sim m_\rho^2$, it is obvious that
the momentum expansion converges slowly at vector meson energy
scale. In the other words, the leading order contribution is not
dominant, so that it is very incomplete to study vector meson
physics only up to the leading order of momentum expansion. In
this paper, we show that the high order terms of momentum
expansion indeed yield very large contribution for vector meson
physics. It implies that a well-defined model including vector
meson resonances must be available to evaluate high order
contribution of momentum expansion.

Some standard methods, such as heat kernel method, are difficult
to capture high order contribution of momentum expansion. In this
paper, therefore, we propose the proper vertex expansion method
to deduce effective action. The resulting effective action
includes all order information of momentum expansion of vector
mesons. Hence, in this paper, we have achieved a complete study on
the chiral expansion theory of EMG model at vector meson energy
scale.

It is crucial and challenging to prove the unitarity of $S$-matrix
for low energy effective theories of QCD. This problem has been
solved satisfactorily in this paper. The unitarity has been
proved order by order in $N_c^{-1}$ expansion. It yields a strong
constrain for all effective theories of low energy QCD, that the
transition amplitude must be real at the leading order of
$N_c^{-1}$ expansion. This constrain can be used to examine
whether an effective theory is eligible or not.

In this paper, we study a number of phenomenological processes
successfully. It should be pointed out that our studies are
different from one in all previous literature, since high order
contributions of momentum expansion and $N_c^{-1}$ expansion are
included in ours consistently. Some new features are revealed in
this papers: 1) For wide resonances, the mass parameter in its
propagator is in general different from its physical mass.
However, the difference between them is model-dependent. So that
the masses of resonances enter effective lagrangian only as a
parameters. Their physical masses should be predicted by relevant
scattering processes. 2) In $\omega\rightarrow\pi^+\pi^-$ decay,
a nonresonant background is allowed, but the contribution from
this back ground is very small. 3) There are three nontrivial
phases in pion form factor (fig. 8). All of these phases are
generated by pion loops and are momentum-dependent.

Up to the next to leading order of $N_c^{-1}$ expansion, all results in
EMG model are factorized by $F_\pi$, $B_0$, $m_{_V}$, $g_{_A}$=(0.75,
$\beta$ decay of neutron), $g=(\pi^{-1}$, KSRF (I) sum rule),
$m$=(480MeV, chiral coupling constant at $O(p^4)$), $\lambda$=(0.54,
Zweig rule) and masses of pseudoscalar mesons. So that EMG model provides
powerful theoretical predictions on low energy meson physics. Furthermore,
since strange quark mass has been extracted from pseudoscalar meson
spectrum, the calculation of this paper can be easily extend to cases
including $K^*(892)$ and $\phi(1020)$. These studies will be found
elsewhere.

\begin{center}
{\bf ACKNOWLEDGMENTS}
\end{center}
This work is partially supported by NSF of China through C. N.
Yang and the Grant LWTZ-1298 of Chinese Academy of Science.

\end{document}